\documentclass[a4paper,12pt]{article}
\usepackage{epsfig}
\usepackage{graphicx}
\usepackage{dcolumn}
\usepackage{bm}
\usepackage{longtable}
\usepackage{amssymb}
\usepackage{amsfonts}
\usepackage[margin=1.0in]{geometry}
\usepackage{amsmath}


\usepackage{slashed}

\def\Re{\mathop{\rm Re}\nolimits}
\def\Im{\mathop{\rm Im}\nolimits}

\def\a{\alpha}

\makeatletter                             
\@addtoreset{equation}{section}                   
\makeatother                              
\usepackage{hyperref}

\begin{document}
\begin{titlepage}
\begin{flushright}
DISIT-2013\\
CERN-PH-TH/2013-041
\par\end{flushright}
\vskip 1.5cm
\begin{center}
\textbf{\huge \bf Entropy Current Formalism} \vspace{.5cm} \\
\textbf{\huge \bf for Supersymmetric Theories}
\textbf{\vspace{2cm}}\\

{\Large L. Andrianopoli$^{~a, c,}$\footnote{laura.andrianopoli@polito.it}, R. D'Auria$^{~a, c,}$\footnote{riccardo.dauria@polito.it},$\ $ \\ \vspace{.5cm}
P.~A.~Grassi$^{~b, d, e,}$\footnote{pgrassi@mfn.unipmn.it},  and
M.~Trigiante$^{~a, c,}$\footnote{mario.trigiante@polito.it}}
\vspace{.5cm}
\begin{center}
 {a) { \it DISAT, Politecnico diTorino, }}\\
 {{
 Corso Duca degli Abruzzi 24,I-10129, Turin, Italy
 }}
\\ \vspace{.2cm}
 {b) { \it DISIT, Universit\`{a} del Piemonte Orientale,
}}\\
{{ \it via T. Michel, 11, Alessandria, 15120, Italy, }}
\\ \vspace{.2cm}
{c) { \it INFN - Sezione di Torino,}}
 \\ \vspace{.2cm}
{d) { \it INFN - Gruppo Collegato di Alessandria - Sezione di Torino,}}
\end{center}
{e) { \it
PH-TH Department, CERN,\\
CH-1211 Geneva 23, Switzerland.}}

\par\end{center}
\vfill{}

\begin{abstract}
{\vspace{.3cm} \noindent The recent developments in fluid/gravity correspondence give a new impulse to the study
of fluid dynamics of supersymmetric theories. In that respect, the entropy current formalism requires some
modifications in order to be adapted to supersymmetric theories and supergravities. We formulate a new entropy
current in superspace with the properties: 1) it is
conserved off-shell for non dissipative fluids, 2) it is invariant under rigid supersymmetry transformations 3)
it is covariantly closed in local supersymmetric theories 4) it reduces to its bosonic expression on space-time.
}

\end{abstract}
\vfill{}
\vspace{1.5cm}
\end{titlepage}

\vfill
\eject

\tableofcontents
\newpage
\setcounter{footnote}{0}


\section{Introduction}\label{sec1}

Recent developments in fluid/gravity correspondence \cite{Bhattacharyya:2008jc,Rangamani:2009xk,Hubeny:2011hd} motivate a
deeper analysis of the fluid dynamics in the context of supersymmetric theories and of supergravity. In the
present work, we take a first step toward that extension by analyzing the definition of the entropy current for
non dissipative fluids (see for example
\cite{Dubovsky:2005xd,Endlich:2010hf,Dubovsky:2011sj,Dubovsky:2011sk,Nicolis:2011cs}) and by providing its
supersymmetric generalization.\footnote{For earlier works on supersymmetric description of fluids see for instance \cite{Hoyos:2012dh}}
The starting point is a convenient formulation of the fluid dynamics
in terms of the comoving coordinates of the  fluid (see
\cite{Brown:1992kc,Jackiw:2004nm}). The Eulerian description in
terms of spacetime-dependent quantities is replaced by a new set
of comoving coordinates $\phi^I$  (with $I = 1, \dots, d$, $d$ being
the space dimensions) which are spacetime fields. In terms of
those,  in the case of
non-dissipative fluids, one can easily write down a Lagrangian whose field equations are the relativistic generalization of the  well-known
Navier-Stokes equations. One can also easily define several
interesting thermodynamical quantities such as the entropy, the
energy density, chemical potentials and so on. This formalism
permits also a direct verification of Maxwell equations for
thermodynamics. Finally, all techniques of quantum field theory can
be used to investigate the quantum properties of fluids (see for
example \cite{Endlich:2010hf}).

Recently a series of interesting papers
\cite{Loganayagam:2008is,Dubovsky:2011sj, Bhattacharya:2012zx}
appeared on the subject by exploring  the fluid dynamics from
the point of view of comoving coordinates and discussing the role of
the entropy current in that context. In particular they claim that
the entropy current of a given system must have the following
properties: 1) It is dual to a $d$-form in $(d+1)$-space-time
dimensions; 2)  It is conserved off-shell. It is easy to show that
the expression
$$
J^{(1)} ={}^*\left(d\phi^1\wedge \dots \wedge d\phi^d\right)\,,
$$
where  the star symbol is the Hodge-dual star operator in $d+1$ dimensions, has the correct properties. In
addition, it cannot be written as a d-exact expression since the comoving coordinates $\phi^I$ are not globally
defined. The entropy density can be computed by considering the Hodge dual of $J^{(1)}$.  In papers
\cite{Dubovsky:2011sj,Nicolis:2011cs,Dubovsky:2011sk}, this formalism has been applied to normal fluids as well
as to superfluids and, there, all quantities are computed in terms of the comoving coordinates and of one
additional degree of freedom $\psi$. A new symmetry has been advocated in order to describe the supefluid in a
suitable phase and the spectrum of waves in that fluid have been taken into account. We briefly review that
formalism in Section \ref{sec2} in order to set up the stage for our developments.

The next step is to provide a supersymmetric extension. Since the coordinates $\phi^I$ represent a set of
comoving coordinates of the fluid, it is natural to introduce a set of anticommuting coodinates $\theta^\a$ for
describing the fluid fermionic degrees of freedom (see \cite{Jackiw:2004nm}). Using the analogy with the
Green-Schwarz superstring and with the supermembrane we define a supersymmetric 1-form $\Pi^I$ replacing the
1-form $d\phi^I$ of the bosonic theory. In that context, we discuss the generalization of the action for
supersymmetric fluids with all symmetries.

Finally, we can provide the supersymmetric extension of the entropy
current. We have to recall that the entropy current is associated
with volume preserving diffeomorphisms  and therefore the
supersymmetric extension must play a similar role of volume
preserving superdiffeomorphisms. In that case the form is an
\emph{integral form }(see for example
\cite{integ,mare,Voronov2,bernstein,Catenacci:2010cs,Witten:2012bg} and references therein) whose
complete expression on the supermanifold ${\cal M}^{(d+1|m)}$ is
$$
{J^{(d|m)} = \frac{1}{d!}\,\epsilon_{I_1 \dots I_d}  \Pi^{I_1}
\wedge \dots \wedge \Pi^{I_d}\, \theta^m \cdots \theta^1\,\wedge \delta(d\theta^{1}) \wedge
\dots \wedge \delta(d\theta^{m})}\,,
$$
$d$ being the spatial dimensions and $m$ the dimension of the spinor representation, that is the number of
fermionic coordinates. \footnote{For the sake of clarity, we  assume here and in the following that the
$\theta^\alpha$ are Majorana spinors (as it is in four dimensions) or Majorana-Weyl spinors. The case of Dirac
or pseudo-Majorana spinors (as it happens e.g. in D=5 supersymmetric theories) can be dealt with in an
analogous way.} This expression transforms as a Berezinian under superdiffeomorphisms and the Dirac delta of
1-superforms $d\theta^\a$ is a symbol that has the usual properties of distributions as explained in
\cite{Catenacci:2010cs}. The Dirac delta-functions of 1-forms $d\theta^\a$ can be understood by assuming that
the fermionic 1-forms are indeed commuting quantities and therefore it becomes pivotal to define an integration
measure in this space. One way -- although it is  not the only one -- is to use  the atomic measure given by the
distributional Dirac delta. The main property of that distribution is  locality which plays an important role in
our construction.

Given the new formula for the entropy current $J$,  we can compute
the entropy density $s$ and we discuss some implications.
As an important application, we generalize our construction to supergravity.

The paper is organized as follows:
\\ In Section 2 we review the Lagrangian approach for fluid dynamics and introduce the entropy current.\\ In
Section 3 we set up the stage for the supersymmetric extension of the bosonic theory and, as a warm-up exercise,  we discuss the symmetries and equations of motion of a supersymmetric  effective theory for fluid
dynamics in a 1+1 dimensional model.
We then extend the considerations  to a general $d+1$ dimensional Lagrangian.
\\ In Section 4 we give a
general expression for the entropy current and entropy density in superspace. We will first discuss the
properties of the entropy current for a 1+1 dimensional model, then for a general $d+1$ dimensional theory
and finally we propose a possible expression of it in $N=1$ supergravity. \\The Appendices contain several
technical details.


\section{Comoving Coordinate Formalism}\label{sec2}
In this section we shortly review the Lagrangian approach developed in ref.s
\cite{Brown:1992kc,Dubovsky:2005xd,Dubovsky:2011sj,Nicolis:2011cs}, which is based on the use of the comoving
coordinates of the fluid as fundamental fields,  adopting the same notations as \cite{Dubovsky:2011sj}. Their
approach will be useful for the extension of the formalism to the supersymmetric case.

From a physical point of view one assumes that the hydrodynamics of
a perfect fluid can be formulated as a low energy effective
Lagrangian of massless fields which are thought of as the Goldstone
bosons of a broken symmetry, namely space translations (broken by
the presence of phonons), and is invariant under the symmetry
associated with conserved charges. The effective complete
Lagrangian would be a derivative expansion in terms of the breaking
parameters (mean free path and mean free time). One tries to
determine the low energy Lagrangian by symmetry requirements.

Working, for the sake of generality, in $d+1$ space-time dimensions,
one introduces $d $ scalar fields $\phi^I(x^I,t)$, $I=1,\dots, d $
as Lagrangian comoving coordinates of a fluid element at the point
${x^I}$ at time $t$ such that, at equilibrium, the ground state is
described by $\Phi^I=x^I$ and requires, in absence of gravitation,
the following symmetries:
\begin{eqnarray}\label{symm}
\delta \phi^I &=& a^I \quad (a^I = \rm const.)\,,\\
\phi^I&\rightarrow &O^I_J\,\phi^J,\,\,\,\,\quad(O^I_J\,\epsilon \,{\rm SO(d)})\,,\label{symm2}\\
\phi^I&\rightarrow &\xi^I(\phi), \quad\quad \det(\partial \xi^I/\partial \phi^J)=1.\label{symm3}
\end{eqnarray}
Furthermore, if there is a conserved charge (particle number, electric charge etc.), then the associated
symmetry
 cannot be described by transformations acting on the fields $\phi^I$, since they   are non compact and they cannot describe particle
number conservation. Therefore one introduces a new field
$\psi(x^I,t)$ which is a phase, that is it transforms under ${\rm
U}(1)$ as follows
\begin{equation}\label{psitrans}
  \psi\rightarrow  \psi + c, \quad (c= \rm const.).
\end{equation}
 Finally one must take into account that the particle number is comoving with the
fluid, giving rise to a (matter) conserved current
\begin{equation}\label{number}
    \partial_\mu j^\mu= 0,
\end{equation}
where
\begin{equation}\label{number1}
    j^\mu= n\,u^\mu\,, \quad u^2 = -1\,,
\end{equation}
$n$ being the particle number density and $u^\mu$ the fluid four-velocity defined below.
Moreover, if the charge flows with the fluid, charge conservation is obeyed
separately by each volume element. This means that the charge conservation is not affected by an arbitrary
comoving position-dependent transformation
\begin{equation}\label{extsym}
   \psi\rightarrow  \psi + f(\phi^I)
\end{equation}
$f $ being an arbitrary function. This extra symmetry requirement on the Lagrangian is dubbed
\emph{chemical-shift symmetry}.

From these premises the authors of \cite{Dubovsky:2011sj}  construct
the low energy Lagrangian respecting the above symmetries. At lowest
order the Lagrangian will depend on the first derivatives of the
fields through invariants respecting the symmetries (\ref{symm})-
(\ref{psitrans}) and (\ref{extsym}) :
\begin{equation}\label{leg}
   \mathcal L = \mathcal L (\partial \phi^I,\partial \psi).
\end{equation}
For this purpose one introduces the following current which respects the symmetries (\ref{symm})-(\ref{symm3}):
\begin{equation}\label{correntona}
    J^\mu=\frac{1}{d!}\, \epsilon^{\mu,\nu_1,\dots,\nu_d}
     \epsilon_{I_1,\dots,I_d} \partial_{\nu_1}\phi^{I_1}\dots \partial_{\nu_d}\phi^{I_d}\,,
 \end{equation}
 and enjoys  the important property that its projection along the comoving coordinates does not change:
\begin{equation}\label{enjoy}
   J^\mu\,\partial_{\mu}\phi^{I}=0.
\end{equation}
Introducing the Hodge-dual star operator in $d+1$ dimensions, this is equivalent to saying that the spatial
$d$-form current

\begin{equation}\label{J0}
  J^{(d)}= -{}^*J^{(1)}=\frac{1}{d!}\,\epsilon_{I_1\dots I_d} d\phi^{I_1}{\wedge} \dots {\wedge} d\phi^{I_d}
\end{equation}
where
\begin{equation}\label{J1}
    J^{(1)}=\frac{1}{d!}\, \epsilon_{\mu \nu_1 \dots \nu_{d}}\,\epsilon_{I_1\dots I_d}\partial^{\nu_1}\,\phi^{I_1}
    \dots \partial^{\nu_d}\,\phi^{I_d}\,dx^\mu =
  (-1)^d \,\,{}^*\left(\frac{1}{d!}\,\epsilon_{I_1\dots I_d} d\phi^{I_1}{\wedge} \dots {\wedge} d\phi^{I_d}\right)\,,
\end{equation}
 is closed identically, that is it is locally an exact form.
Hence it is natural to define the fluid four-velocity as aligned with $J^\mu$:
\begin{equation}\label{align}
    J^\mu=b\,u^\mu \rightarrow b=\sqrt{-J^\mu\,J_\mu}=\sqrt{{\rm det}( {B}^{IJ})}\,,
\end{equation}
where $ {B}^{IJ}\equiv \partial_\mu \phi^I \partial^\mu \phi^J$.
From a physical point of view, the property of $J^\mu$ to be
identically closed identifies it as the entropy current of the
perfect fluid, so that $b=s$,  $s$ being  the entropy density. Using
the entropy current $J^\mu$ one finds that, by virtue of eq.
(\ref{enjoy}), the quantity $J^\mu \partial_\mu\psi$ is invariant
under (\ref{extsym}).

Summarizing, a low energy action invariant under (\ref{symm})- (\ref{psitrans}) and (\ref{extsym}), can depend
on $\phi^I$ and $\psi$ only through $J^\mu$ and $J^\mu
\partial_\mu\psi$, and, being a Poincar\'{e} invariant, it can be written as follows:
\begin{equation}\label{symmlagr}
   \mathcal{S}=\int d^4\,x F(b,y)\,,
\end{equation}
where $y$ is
\begin{equation}\label{yy}
   y= u^\mu \partial_\mu \psi=\frac{ J^\mu\,\partial_{\mu}\psi}{b}\,.
\end{equation}
Computing the Noether current associated with the symmetry (\ref{psitrans}) one derives
\begin{equation}\label{noether}
   j^\mu=F_y\,u^\mu \rightarrow  F_y\equiv n,
\end{equation}
which identifies $n$ as the particle number density. The Noether currents associated with the infinite symmetry
(\ref{extsym}) are
\begin{equation}\label{noether1}
  j^\mu_{(f)}=F_y\,u^\mu f(\phi^I),
\end{equation}
 and these currents are also conserved by virtue of the $j^\mu$-conservation.

 By coupling (\ref{symmlagr}) to worldvolume gravity we can obtain the energy-momentum tensor by taking,
 as usual, the derivative with respect to a background metric:
\begin{equation}\label{finaltmunu}
   T_{\mu\nu}=\left(y\,F_y-b\,F_b\right)u_\mu\,u_\nu +\eta_{\mu\nu}\left(F-b\,F_b\right)\,.
\end{equation}
On the other hand, from classical fluid-dynamics, we also have
\begin{equation}\label{finaltmunucl}
   T_{\mu\nu}=\left(p+\rho\right)u_\mu\,u_\nu +\eta_{\mu\nu}p\,,
\end{equation}
from which we identify the pressure and density
\begin{equation}\label{ossia}
   \rho =y\,F_y-F\equiv y\,n-F\,\,\,,\,\,\,
p= F-b\,F_b\,.
\end{equation}
From the derivation of the energy-momentum tensor one can easily
obtain the entropy density, the temperature, the Maxwell equations
and so on (see \cite{Dubovsky:2011sj} for a complete review).
 In particular, it
turns out that the quantity $y$ defined in eq. (\ref{yy}) coincides
with the chemical potential  $\mu$. To see this, it suffices to
compare the first principle
\begin{equation}\label{densenergy1}
  p+\rho=T\,s +\mu\,n\,,
\end{equation}
 with (\ref{ossia}). Using $F_y=n$ and $b=s$, we then find
\begin{equation}\label{comp}
   \frac{\partial F}{\partial s}=-T \,\,,\,\,\, y =\mu\,.
\end{equation}
We conclude that the Lagrangian density is a function of $s$ and
$\mu$

\begin{equation}\label{concl}
    F=F(s,\mu)\,.
\end{equation}

Let us remark that the Lagrangian used in this setting
does not allow at first sight for the presence of a kinetic term for the dynamical field $\psi$, namely $X=\partial_\mu \psi
\partial^\mu \psi$. In fact we could  also consider, besides $X$, further Poincar\'e invariants of the form $Z^I=\partial_\mu \psi
\partial^\mu \phi^I$. However, it can be proven that the quantities $X,\,Z^I$, together with $B^{IJ}$,  are not independent of $y$ since the following relation holds:
\begin{equation}
y^2 = - \partial_\mu \psi
\partial^\mu \psi + \partial_\mu \psi
\partial^\mu \phi^I B^{-1}_{IJ}\partial_\nu \psi
\partial^\nu \phi^J\,.
\end{equation}
Therefore a dependence of the Lagrangian on $X$ is somewhat implicit in $y^2$.\par
A Lagrangian exclusively depending on $X$, i.e. of the form $F(X)$,  has been considered, for instance,
 in \cite{Nicolis:2011cs}, to describe superfluids at $T=0$.  The use of the variables $X,Z^I$,
  even though redundant for ordinary fluids, can be useful in order to describe superfluids as a spontaneously broken phase
 of a field theory with chemical-shift symmetry invariance.  This idea  is  elaborated in \cite{noi}.

\section{Supersymmetric Extension. Lagrangian and Equations of Motion}
In order to generalize to the supersymmetric case the lagrangian formalism and the definition of the entropy
current reviewed in the previous section let us first set up our formalism.\\
 A basis of 1-superforms in a general
\emph{rigid} $(d+1 {\mid} m)$-superspace can be given in terms of the supervielbein $\{\Pi^a,\Psi^\alpha \}$ by:

\begin{equation}\label{defviel}
\Pi^a= d\phi^a + \frac i2\bar\theta \Gamma^a d\theta\,,\quad \Psi^\alpha = d\theta^\alpha
\end{equation}
where $a=(0,I),(I= 1,\cdots d)$ and $\alpha=(1,\cdots m)$ run over the bosonic and fermionic directions of
superspace respectively. Here $\Gamma^a$ are the Clifford algebra $\Gamma$-matrices in $d+1$-dimensions, while
$\theta $, and $d\theta$ denote the matrix form of the \emph{Majorana} spinors in the $m=2^{d/2}$-dimensional
spinor representation of ${\rm SO}(d,1)$. \footnote{For  \emph{Majorana-Weyl} spinors, the dimension of the
representation is instead $m=2^{(d-1)/2}$.} In particular $\bar{\theta}\equiv\theta^\dagger \Gamma^0=\theta^T\,C$, $C=(C_{\alpha\beta})$ being the charge-conjugation matrix. The space-like fields $\phi^I(x,t)$ can be taken as the comoving
coordinate fields of the bosonic theory, while the spinors $\theta^\alpha (x,t)$ are the fermionic coordinates
of superspace and we added a time-like bosonic field $\phi^0$ to complete the superspace (see also \cite{Jackiw:2004nm})).\\
Together with the supersymmetric extension of $d\phi^I$ we also introduce a 1-form $\Omega$, representing the
supersymmetric extension of the chemical shift field-strength $d\psi$:

\begin{equation}\label{omega}
    \Omega = d \psi +i\bar \tau \, d\theta\,.
\end{equation}
$\tau$ being a new Majorana spinor.

The tangent vectors
\begin{equation}\label{defD}
     \partial_a =\frac{\partial}{\partial{\phi^a}}\;\,,\quad\quad  D_\alpha= \frac{\partial}{\partial \theta^\alpha}
     +\frac i 2\,(\bar \theta\,\Gamma^a)_\alpha\, \partial_a
\end{equation}
are dual to the supervielbein and the related supersymmetry transformations are:
\begin{equation}\label{SN22d}
\delta_{\bar\epsilon  D} \phi^a = \frac i 2\bar\epsilon \Gamma^a \theta\,, \quad \delta_{\bar\epsilon  D} \theta = \epsilon\,,
\end{equation}
while  the supervielbein transforms as\footnote{Note that the supervielbein $\{\Pi^a,\psi^\alpha \}$ are left
invariant under the Killing vectors transformations $\bar Q_\alpha $
 generators of the supersymmetry algebra, namely

\begin{align}\label{SN8}
&\delta_{\bar\epsilon  Q} \phi^a = -\frac i 2\bar\epsilon \gamma^a \theta\,, \quad \delta_{\bar\epsilon  D} \theta = \epsilon\\
& \delta_{\bar \epsilon Q} \Pi= \delta_{ \bar\epsilon  Q}d\,\theta =0
\end{align}
where

\begin{align}
& Q_\alpha=\frac{\partial}{\partial \theta^\alpha} -\frac i 2\,(\bar \theta\,\gamma^a)_\alpha\, \partial_a \longrightarrow
 \{  Q_\alpha,\, D_\beta\}=0
\end{align}
are the fermionic generators of the superPoincar\'{e} algebra (anti)commuting with the tangent vectors $D_\alpha$.}
\begin{equation}
\label{SN222d}
\delta_{\bar \epsilon \,D} \Pi^a =  i \bar\epsilon \Gamma^a d\theta\,, \quad \delta_{\bar \epsilon D} \psi = d\epsilon=0\,.
\end{equation}

We further assume that the phase $\psi(x)$ and its supersymmetric partner $\tau(x)$ are invariant under the
rigid supersymmetry generated by the Killing vector $\vec \epsilon.$ We may, however, extend the chemical shift
``internal" symmetry to superspace performing the following superdiffeomorphism on $\psi$:

\begin{equation}\label{superchem}
   \delta \psi = f(\phi,\theta)
\end{equation}
with $f(\phi,\theta)$ arbitrary superfield. Assuming $\delta \tau = Df$, the 1-form $\Omega$ acquires the
following chemical shift transformation:

\begin{equation}\label{deltaomega}
   \delta \Omega =\frac{\partial f}{\partial \phi^I} \Pi^I.
\end{equation}

 In order to catch the relevant points of the supersymmetric generalization of the
bosonic theory in the simplest way, it is convenient to first restrict ourselves to the $1+1$ dimensional case,
that will be dealt with in the next subsection, postponing its extension to $(d+1)$ space-time dimensions to the
following subsection.

\subsection{Supersymmetric Effective Theory in Two Space-Time Dimensions
 } \label{sec3}

In 1+1-dimensions we have just one comoving coordinate $\phi(x,t)$ and the embedding superspace has two bosonic
and one fermionic dimensions.\footnote{Recall that in 1+1-dimensions the spinors are Majorana -Weyl so that
they have just one component.} Therefore eq.s (\ref{defviel}) and (\ref{omega}) imply:

\begin{equation}\label{SN1bis} \Pi = d \phi + \theta \, d\theta\,, \quad \Psi=d\theta,
\quad\quad
\Omega = d \psi + \tau \, d\theta\,,
\end{equation}
where we have denoted by $\phi$ the spatial component of $\phi^a$ in two dimensions.
The supersymmetry transformations are generated by
\begin{equation}\label{2dD}
    D = \partial_\theta + \theta \partial _\phi\,,
\end{equation}
which satisfies $D^2 = - \partial_\phi$. Setting  $\epsilon D\equiv \vec \epsilon$, the supersymmetry
transformations are now:
\begin{equation}\label{SN2}
\delta_{\vec \epsilon} \phi = \epsilon  \theta\,, \quad \delta_{\vec \epsilon} \theta = \epsilon\,,
\end{equation}
while  the spatial component of the bosonic vielbein transforms as
\begin{equation}
\label{SN22}
\delta_{\vec \epsilon} \Pi =  2 \epsilon  d\theta\,, \quad \delta_{\vec \epsilon} d\theta = d\epsilon=0\,,
\end{equation}
since the spinor parameter $\epsilon$ is constant. Furthemore

\begin{equation}\label{susychem}
   \delta_{\vec \epsilon} \psi=\delta_{\vec \epsilon} \tau =0\,.
\end{equation}

In terms of these variables, we can build the following quantities
\begin{eqnarray}\label{SN4}
&& B= -\Pi \wedge  {}^\star  \Pi = \Pi_\mu g^{\mu\nu} \Pi_\nu d^2x=\hat{B}\,d^2x\,,\nonumber\\
&& Y = \Omega \wedge \Pi = \epsilon^{\mu\nu} \Omega_\mu \Pi_\nu d^2x=\hat{Y}\,d^2 x\,,
\end{eqnarray}
where we denoted by the same letter, though with a hat on the top, the corresponding quantity modulo the volume
form, that is $\hat B=  \Pi_\mu \eta^{\mu\nu} \Pi_\nu$, $\hat Y=  \Omega_\mu \epsilon^{\mu\nu} \Pi_\nu$,
$b=\sqrt{\hat B}$.

The variation  of the Poincar\'e-invariant superfields under a generic variation of $\phi$ and of $\psi$ is
\begin{eqnarray}\label{SN5}
&& \delta B = -2 \Pi \wedge {}^\star  d \delta \phi \,,  \nonumber \\
&& \delta Y = - \Pi  \wedge d \delta \psi  + \Omega \wedge  d \delta \phi  \,. \nonumber \\
\end{eqnarray}
Therefore, if the action is given by
\begin{equation}
 \mathcal{S}= \int d^2x F[b,Y]\,,\label{susys}
 \end{equation}
as an integral of a local functional, we get the equations of motion

\begin{eqnarray}\label{SN6}
&& d \Big[ b^{-1}\, F_b {}^\star  \Pi + F_Y \Omega  \Big] =d{}^\star  J_\phi=0\,, \\
&& d \Big[  F_Y \Pi  \Big] =d{}^\star  j=0\,,
\end{eqnarray}
where $ F_b = \partial F / \partial b\,,  F_Y = \partial F / \partial Y$, and we have defined the 1-forms:
\begin{equation}
J_\phi=b^{-1}\, F_b  \Pi + F_Y {}^\star \Omega \,\,;\,\,\,j=   F_Y {}^\star \Pi \,,
\end{equation}
which are the Noether currents associated with the shift symmetries $\phi\rightarrow \phi+c',\,\,\psi\rightarrow \psi+c$.
 The variations for $\theta$ and $\tau$ are
\begin{eqnarray}\label{SN7}
&& \delta B =- 2 \Pi \wedge {}^\star  (\delta \theta d\theta + \theta d \delta \theta)  \,,  \nonumber \\
&& \delta Y = - \Pi  \wedge (\delta \tau d\theta + \tau d \delta \theta)  +
\Omega \wedge (\delta \theta d\theta + \theta d \delta \theta)  \,, \nonumber \\
\end{eqnarray}
and the corresponding equations of motion are
\begin{eqnarray}\label{SN8b}
&&
2 {}^\star  J_\phi\wedge d\theta
+  d\tau \wedge
{}^\star  j  =0\,, \nonumber \\
&&
{}^\star  j  \wedge d\theta = 0\,,
\end{eqnarray}

We can also compute the supercurrent and the current associated with
the chemical shift symmetry. Recalling that the supercurrent is obtained by variation with respect to $d\theta$, we find
\begin{equation}\label{SN9A}
j_S =  -2\,\theta\, {}^\star  J_\phi
+  \tau\, {}^\star  j\,,
\end{equation}
which, by the equations of motion, enjoys the property: $d {}^\star  j_S =0$.

We now show that the action functional is \emph{invariant under supersymmetry.} Indeed, if we  restrict the
variation of the action (\ref{susys}) to the supersymmetry transformations  (\ref{SN2}), (\ref{SN22}) and
(\ref{susychem}) we find

\begin{equation}\label{susyaction}
    \delta_{\vec \epsilon} \mathcal{S}= 2\int [{}^\star  J_\phi\wedge d(\epsilon\theta)]
\end{equation}
so that, by partial integration and use of the equation of motion, it follows

\begin{equation}\label{susyaction1}
    \delta_{\vec \epsilon} \mathcal{S}= 0.
\end{equation}
This is of course a major result of our approach.

Concerning the chemical shift symmetry, we consider the possibility of constant symmetry, namely $f = a +
\omega \theta$ where $a$ and $\omega$ are communting and anticommuting parameters of constant type,
respectively. Therefore, it is easy by Noether method to compute the following two currents
\begin{eqnarray}\label{SN10}
J^{(a)} = j\,, \quad\quad
J^{(\omega)} =j\,\theta \,, \quad\quad
\end{eqnarray}
which can be obviously cast into a supermultiplet. Notice that $d {}^\star  J^{(a)} =0$ as follows from the
second equation of (\ref{SN6}), while $d{}^\star  J^{(\omega)} =0$ follows from the second equation of
(\ref{SN8b}).\\

\subsection{The General $d+1$ Dimensional Case}\label{sec4}
In this section  we generalize the lagrangian  given in the previous
section in the $1+1$-dimensional case to superspace with $(d+1)$-space-time dimensions and $m$ fermionic directions.
\par
We generalize  equations (\ref{SN4}) by defining
 the following quantities:
\begin{eqnarray}\label{SSN4}
&& B^{IJ}=- \Pi^I \wedge  {}^\star  \Pi ^J=\Pi^I_\mu\Pi ^{J\mu}\, d^{d+1} x=\hat{B}^{IJ}\,d^{d+1} x\,,
\nonumber\\
&& Y = \Omega \wedge \Pi ^1 \cdots \wedge \Pi^d=
\frac 1{d!}\epsilon^{\mu\nu_1\cdots\nu_d}\epsilon^{I_1\cdots I_d} \Omega_\mu \Pi^{I_1}_{\nu_1}\cdots \Pi^{I_d}_{\nu_d}\ d^{d+1}x=\hat{Y}\,d^{d+1}x\,, \nonumber\\
\end{eqnarray}
where we have denoted by the same letter, though with a hat on the top, the corresponding factor multiplying
$d+1$-dimensional volume form. Note in particular that \cite{Dubovsky:2011sj}:
\begin{equation}
\hat b  =
\sqrt{\det \hat B^{IJ}}\label{b}\,,\end{equation} and \begin{equation}
\left.\hat{b}\right\vert_{\theta=0}=s\,\,;\,\,\,\left.\frac{\hat{Y}}{\hat{b}}\right\vert_{\theta=0}=y\,.
\end{equation}
where $b=s$ and $y=\mu$ are the entropy density and the chemical potential as defined in Section 2.

 On the basis of the previous discussion the action generalizing (\ref{susys}) will be  written as the following integral of a local functional
 \begin{equation}\label{susysd}
   \mathcal{S}= \int d^{d+1}x F[{\hat b}, \hat Y].
 \end{equation}
The results of the $1+1$-dimensional case are then easily generalized as follows. The variation  of the
Poincar\'e-invariant superfields under a generic variation of $\phi^I$ and of $\psi$ is
\begin{eqnarray}\label{SSN5}
&& \delta B^{IJ}  =- 2 \Pi ^I\wedge {}^\star  d \delta \phi^J \,,  \nonumber \\
&& \delta Y =d \delta \psi \wedge \Pi ^1 \wedge \cdots \wedge \Pi^d   +
 \frac {1} {(d-1)!}\epsilon^{I_1\cdots I_d} \Omega\wedge  \Pi^{I_1}\wedge \cdots\wedge \Pi^{I_{d-1}}\wedge
 d \delta \phi ^{I_d} \,, \nonumber \\
\end{eqnarray}
and the following equations of motion are obtained:
\begin{align}\label{SSN6}
&d{}^\star  J_I^{(1)}= d\Big[ {\hat b}\,F_b \hat{B}^{-1} _{IJ} {}^\star  \Pi^J +\frac {1 } {(d-1)!}\epsilon^{I J_1\cdots J_{d-1} } F_Y\Omega\wedge  \Pi^{J_1}\wedge
 \cdots\wedge \Pi^{J_{d-1}} \,, \\
& d{}^\star  j^{(1)}= d \Big[   F_Y \Pi ^1 \wedge\cdots \wedge\Pi^d\Big] =0\,,
\end{align}
where $ F_b = \partial F / \partial {\hat b}\,, F_Y = \partial F / \partial \hat{Y}$ and we have introduced the
two currents:
\begin{align}
\label{deltaphi}J_I^{(1)}&= {\hat b}\,F_b \hat{B}^{-1} _{IJ}\Pi^J + \frac {1 } {(d-1)!}
\Omega_\mu\epsilon^{\mu\nu\mu_1\dots \mu_{d-1}}\epsilon^{I J_1\cdots J_{d-1} } F_Y
 \Pi^{J_1}_{\mu_1}
 \cdots \Pi^{J_{d-1}}_{\mu_{d-1}}dx_\nu \,,\\
\label{deltaphi2}j^{(1)}&=  - F_Y\,\epsilon^{\mu\mu_1\dots\mu_d}dx_\mu \Pi^1_{\mu_1}\dots \Pi^d_{\mu_d}=(-)^{d+1}\, F_Y\,{}^\star J^{(d)}\,,
\end{align}
where
\begin{equation}\label{firstdef}
   J^{(d)}=\Pi^1\wedge \cdots\wedge \Pi^d\equiv \frac{1}{d!}\epsilon_{I_1 \dots I_d}  \Pi^{I_1}  \wedge \dots \wedge \Pi^{I_d}.
\end{equation}
 It is straightforward to verify that the (\ref{deltaphi}) and (\ref{deltaphi2}) are  the Noether currents associated with the
constant translational symmetries $\phi^I\rightarrow \phi^I+c^I$ and $\psi\rightarrow \psi+c$. Furthermore, a
general variation  of the fermionic fields $\theta$ and $\tau$ gives
\begin{align}\label{SSN7}
& \delta B^{IJ} = -i \Pi^I \wedge {}^\star  (\delta\bar \theta \Gamma^J d\theta + \bar\theta\Gamma^J d \delta \theta)  \,,  \nonumber \\
&\delta Y =i\,(\delta \bar\tau d\theta +\bar \tau d \delta \theta)\wedge J^{(d)}  +\frac i 2
 \frac {1 } {(d-1)!}\epsilon_{I_1\cdots I_d} \Omega\wedge  \Pi^{I_1}\wedge \cdots\wedge \Pi^{I_{d-1}}\wedge
 (\delta\bar \theta \Gamma^{I_d} d\theta + \bar\theta\Gamma^{I_d} d \delta \theta) \,,
 \end{align}
so that the corresponding \emph{fermionic} equations of motion are
\begin{align}
J_I^{(1)\mu}\,\Gamma^I\partial_\mu\theta+\eta_C\,j^\mu\,\partial_\mu \tau=0\,,
\end{align}
where $\eta_C$ is the sign appearing in the relation $\bar{\tau}d\theta=\eta_C\,\bar{d\theta}\tau$ and depends
on the property of the charge-conjugation matrix in $(d+1)$-dimensions.\par
Finally the equation of motion obtained by varying $\tau$ reads:
\begin{equation}
J^{(d)}\wedge d\theta=0\,.
\end{equation}

\subsubsection{The Energy-Momentum Tensor}\label{enmom}
In equation (\ref{b}) we have seen that the generalization of the entropy density $b\equiv s$
\begin{equation}\label{genentropy}
 s(x)=   s[\partial_\mu\phi^I]= \sqrt{\det \partial_\mu\phi^I\, \partial^\mu\phi^J}
 \end{equation}
 to a  supersymmetric setting is simply obtained by replacing the purely spatial rigid vielbein $d\phi^I$ with its
 supersymmetric version $\Pi^I(x,\theta)=d\phi^I +\frac{\rm i}{2}{\overline\theta \Gamma^I}d\theta$,  so that the entropy density superfield
 is given by
 \begin{equation}\label{superentro}
\hat s(x)= \hat s\left[\partial_\mu\phi^I(x)\,,\partial_\mu \theta^\alpha(x)\right]=  \sqrt{\det  \Pi^I_\mu\, \Pi^{J \mu}}.
 \end{equation}
In an analogous way we generalize the bosonic variable of Section 2,  $y=(J^\mu /s)\, \partial_\mu \psi$, to
\begin{equation}\label{Y}
   \hat y=\frac{J^{\mu (1|0)}}{\hat s}\,\Omega_\mu=\frac{\hat{Y}}{\hat{s}}\,,
\end{equation}
where $J^{(1|0)\mu}$ is the natural extension of the current  $J^{(1)}$ defined in equations (\ref{J0}) and
 (\ref{J1}), namely
\begin{equation}\label{zeropicture}
    J^{(1|0)\mu}=\frac 1{d!} \epsilon^{\mu\nu_1\cdots \nu_d}\epsilon_{I_1 \dots I_d}   \Pi^{I_1}_{\nu_1} \dots  \Pi^{I_d}_{\nu_d}\,.
\end{equation}
the extra index zero being added to comply with the notation of the \emph{integral forms} given in the next
section.\\ The generalization  of the energy-momentum tensor to the supersymmetric theory is then obtained by
varying the action functional of the superfields, $\mathcal{S}$, given by 
(\ref{susysd}):
\begin{equation}\label{action2}
   \mathcal{S}=\int d^{d+1}\,x \sqrt{-g}\,F(\hat s,\hat y)\,,
\end{equation}
with respect to a probe metric $g_{\mu\nu}$. It is straightforward to see that we obtain for the energy-momentum
tensor \emph{formally} the same result as in the non-supersymmetric case, the only difference being the
substitution of $b,y$ with their supersymmetric counterparts $\hat s(x,\theta ), \hat y(x,\theta)$:
\begin{equation}
 T_{\mu\nu}= \left(\hat y \frac{\partial F}{\partial \hat y} - \hat s \frac{\partial F}{\partial \hat s}\right) u_\mu u_\nu +
  \eta_{\mu\nu}\,\left(F- \hat s \frac{\partial F}{\partial \hat s}\right)\,.
 \end{equation}
It follows that also the relation between the thermodynamic functions and the superfields $\hat s(x,\theta ),
\hat y(x,\theta)$ remains formally the same, namely
\begin{eqnarray}\label{action3}
    p&=&F(\hat s,\hat y)-\hat s(x,\theta)F_s\,,\\
    \rho &=&\hat y(x,\theta)\,\hat n-F(\hat s,\hat y)\,,
\end{eqnarray}
where $\hat n(x,\theta)$ is the superfield generalizing the particle number density. We see that the pressure,
the energy density and  the particle number density become superfields whose $\theta=0$ components give the
usual field variables $\rho(x),p(x), n(x)$. The same of course happens for all the other thermodynamical
variables.\\
The generalization of the entropy density given in equation (\ref{superentro}) implies that the current
$J^{(d)}$ defined in equation ({\ref{firstdef}}) should be the natural extension to the supersymmetric case of
the bosonic entropy current \footnote{Here and in the following we will generally call  ``entropy current" the
$d$-form associated (through Hodge-duality) to the entropy-current 1-form.} since its $\theta=0$ component
coincides with equation (\ref{J1}). However, it does not enjoy the important property of being closed, so that
it raises several problems of interpretation. For this reason we shall introduce in the next section, using the
framework of the integral forms, two alternative definitions of entropy current.

\section{The Supersymmetric Entropy Current}

The formalism of the \emph{integral forms} allows us to generalize the notion of entropy  current to a supersymmetry-invariant (modulo a total derivative), closed form defined on a $(d+1|m)$-superspace. Such generalization is however not unique.
We could consider, for instance, the following supersymmetric extension of the
purely bosonic entropy current (\ref{correntona}):
\begin{equation}\label{EC4}
{J^{(d|m)} = \frac{1}{d!}\epsilon_{I_1 \dots I_d}  \Pi^{I_1}  \wedge \dots \wedge
 \Pi^{I_d} \bigwedge_{\alpha=1}^m\delta(d\theta^{\alpha})}=J^{(bos)}\,\bigwedge_{\alpha=1}^m
\delta(d\theta^{\alpha})\,,
\end{equation}
where $J^{(bos)}$ coincides with the $J^{(d)}$ entropy current of the purely bosonic theory, equation (\ref{J0}), as we are going to show below. \\
It has the
following properties:
\begin{enumerate}
\item It is a purely spatial $d$-form on space-time;
 \item It is conserved off-shell;
\item Its zero-picture part, at $\theta=0$, reduces to the bosonic expression in ordinary hydrodynamics;
\item It is invariant under supersymmetry.
\end{enumerate}

The expression (\ref{EC4}) is a so called  \emph{integral form}. Its distinctive feature is the presence of the
distribution $\delta(d\theta)$ (see Appendix \ref{A2}), defined as the Dirac delta-function of the differential
$d\theta$ (recall that
 $d\theta^{\alpha}$ are commuting quantities) and $m$ is the dimension of the spinor representation in
 $(d+1)$-dimensions.

To prove the above properties, it is mandatory to  recall the main properties of integral forms (see Appendix
\ref{A2}).
\par First of all $\delta(d\theta^\alpha)$  enjoys the usual equation $d\theta^\alpha \,\delta(d\theta^\alpha) =
0$. This justifies the alternative expression of $J^{(d|m)}$ in terms of $J^{(bos)}$ given in equation
(\ref{EC4}). In addition, $\delta(d\theta^\alpha)$ carries no form-degree and therefore, multiplying it by any
number of Dirac delta functions

$$
\delta(d\theta^{\alpha_1}),\,\delta(d\theta^{\alpha_2}),\dots
$$
a $d$-form remains a $d$-form. However, we can assign a new quantum number $q$, dubbed  {\it picture number},
which takes into account the number of the Dirac delta-functions $\delta(d\theta^\alpha)$.
 Thus a $p$-form of picture $q$ is denoted by $\omega^{(p|q)}$. Notice that
by using the properties of the Dirac delta-functions it is easy to show that $\delta(d\theta^\alpha) {\wedge}
\delta(d\theta^\beta) = - \delta(d\theta^\beta) {\wedge} \delta(d\theta^\alpha)$; as $d\theta^\alpha$ are
commuting quantities, we have instead $d\theta^\alpha\wedge d\theta^\beta =
d\theta^\beta{\wedge}d\theta^\alpha$. Therefore any integral in superspace of a $p$-form ($p\leqq d+1$) with a
given picture-number $q$ cannot have more than one delta-function of a given differential $d\theta^\alpha$.
\\One can also consider any number $n$ of
 derivatives of a Dirac delta-function $\delta^{(n)}(d\theta)=\frac{d^n\delta(d\theta)}{d(d\theta)^n} $,
 each derivative lowering the degree of the form of one unit. Therefore an additional factor $\delta^{(n)}(d\theta)$
  lowers the form degree of $n$ units. This implies that we can have forms of negative degree.
  In particular one can show the identity $d\theta\,\delta'(d\theta)=-\delta(d\theta)$
  and, in general, $(d\theta)^n\,\delta^{(n)}(d\theta)=(-)^n\,\delta(d\theta)$. Finally, as far as integration is concerned,
  the following formula holds (see eq. (\ref{comE})) :
\begin{equation}
\int   \prod_{\alpha=m}^1\theta^\alpha\bigwedge_{\beta=1}^m \delta(d\theta^{\beta})=1\,.\label{theproperty}
\end{equation}
For a more detailed discussion see Appendix A2.

The entropy current (\ref{EC4}) is exactly invariant under rigid supersymmetry as it can be
easily proven as follows:
\begin{align}
\delta_{\vec{\epsilon}}{ J}^{(d|m)}&=\frac{1}{(d-1)!}\epsilon_{I_1 \dots I_d}  \Pi^{I_1}  \wedge \dots \wedge
 \Pi^{I_{d-1}}\wedge \frac{i}{2}\,\bar{\epsilon}\Gamma^{I_d}d\theta \bigwedge_{\alpha=1}^m \delta(d\theta^{\alpha})=0\,,
\end{align}
where we have used the property
$\delta_{\vec{\epsilon}}\delta(d\theta^{\alpha})=\delta'(d\theta^{\alpha})\,d\epsilon^\alpha=0$, and the fact
that the $d\theta$ in the variation of $\Pi^I$ is annihilated by the integral forms. This current is also
trivially conserved off-shell. \par
Although the zero-picture part of $J^{(d|m)}$ yields the right bosonic entropy current, it would be desirable to relate it to the fluid entropy $S$ in  geometrical way which is intrinsic to the supermanifold: Just as the entropy is the integral of $J^{(bos)}$ over a spatial hypersurface, it would be natural to express, in a supersymmetric context,  the same entropy as the integral over a spatial
$(d|m)$ hypersurface $(S\Sigma)_t^{(d|m)}$ at a given time $t$, of the corresponding supersymmetric current:
\begin{equation}
S(t)=\int_{(S\Sigma)_t^{(d|m)}} J^{(d|m)}\,.
\end{equation}
A drawback of the   definition (\ref{EC4}) is however that the integral of ${ J}^{(d|m)}$
over a super-hypersurface does not reduce to the integral over
a spatial hyper-surface of $J^{(bos)}$. In fact  such an integral would be
zero since:
\begin{equation}
\int_{(S\Sigma)_t^{(d|m)}} J^{(d|m)}=\int_{(S\Sigma)_t^{(d|m)}}J^{(bos)}\,\bigwedge_{\alpha=1}^m
\delta(d\theta^{\alpha})=\int_{(\Sigma)_t^{(d)}}\int_{Berezin}J^{(bos)}=0\,,
\end{equation}
where, for the integration of integral forms,  we have used the prescription in \cite{Catenacci:2010cs}, summarized in Appendix \ref{A2} (see in particular eq. (\ref{comE}).).  One
could still express the entropy in terms of the integral over the whole superspace of a quantity of the form
${ J}^{(d|m)}\wedge {}^*{ J}^{(d|m)}$, by suitably defining the Hodge-star operation in superspace.
Such definition is however subtle and we refrain from dealing with it here.

\vskip 5mm The generalization of the bosonic entropy current to a closed supersymmetry-invariant (modulo a total
derivative) current in superspace is however not unique. Instead of (\ref{EC4}), we could alternatively define a
(super) entropy current in a slightly different fashion as follows:
 \begin{equation}\label{EC4alter}
\boxed{{{\cal J}^{(d|m)} = \frac{1}{d!}\epsilon_{I_1 \dots I_d}  \Pi^{I_1}  \wedge \dots \wedge
 \Pi^{I_d} (\theta)^m \bigwedge_{\alpha=1}^m \delta(d\theta^{\alpha})}}\,,
\end{equation}
where we have used the short-hand notation:
\begin{equation}
(\theta)^m\equiv \prod_{\alpha=m}^1\theta^\alpha\,.
\end{equation}
The quantity ${\cal J}^{(d|m)}$ has the following properties:
\begin{enumerate}
\item It is a purely spatial $d$-form on space-time and a Grassmann density in superspace;
 \item By integration on the fermionic volume element it reduces to the bosonic expression $J^{(bos)}$;
 \item It is conserved off-shell;
\item As it transforms as a total differential under the supersymmetry transformations  (\ref{SN2}),(\ref{SN22}),
its associated conserved charge, \emph{ the entropy, is invariant under supersymmetry.}
\end{enumerate}
In particular the second property allows us to consistently write
\begin{equation}
S(t)=\int_{(S\Sigma)_t^{(d|m)}} {\cal J}^{(d|m)}\,,\label{SrelJ}
\end{equation}
which is the natural relation we were looking for.\par
To prove the above properties,  let us first inspect the simpler, $1+1$-dimensional case, where all the relevant properties
 of the entropy current are already at work, and then the general $d+1$ dimensional case.

\subsection{Two-dimensional case}\label{2dentr}
According to the properties of ${\cal J}^{(d|m)}$ listed above, in the 1+1-dimensional case the entropy current
must satisfy the following requirements:
\begin{enumerate}
\item It is a one-form in the bosonic coordinates and a density in the fermionic sector of superspace.
\item By integration on the fermionic volume element it reduces to the bosonic expression $d\phi$.
\item It is conserved off-shell.
\item It transforms as a total differential under the supersymmetry transformations  (\ref{SN2}),(\ref{SN22}).
\end{enumerate}
The following expression satisfies the requirements:
\begin{equation}\label{EC1}
{\cal J} = \Pi \wedge \theta\,\delta(d\theta)\equiv d\phi \wedge \theta\,\delta(d\theta)\,,
\end{equation}
where $\delta(d\theta)$ is the Dirac delta function of the differential $d\theta$.
 \vskip 0.5 cm

With the definitions introduced above, we can write the entropy current  in (\ref{EC1}) as ${\cal J}^{(1|1)}$.
By construction  the superspace integration of ${\cal J}^{(1|1)}$ reduces to the bosonic
expression $J^{(bos)}= d\phi$, so that the second property is satisfied.
 Moreover we note that ${\cal J}^{(1|1)}
= \Pi \wedge \theta\,\delta(d\theta) = d\phi \wedge \theta\,\delta(d\theta)\equiv J^{(bos)}\wedge
\theta\,\delta(d\theta) $ where we have used $d\theta \wedge \delta(d\theta)=0$. It follows

\begin{equation}\label{requir4}
    d{\cal J}^{(1|1)}= d(J^{(bos)}\wedge
\theta\,\delta(d\theta)) = d\,(d\phi \wedge
\theta\,\delta(d\theta))=- d\phi\wedge d\theta\, \delta(d\theta)=0\,.
\end{equation}
Finally let us show its property under a supersymmetry transformation:

\begin{equation}\label{susytraJ}
   \delta_{\vec \epsilon}{\cal J}^{(1|1)}\equiv \ell_{\vec \epsilon}{\cal J}^{(1|1)}=
   \left(\iota_{\vec \epsilon} d+d\,\iota_{\vec \epsilon}
    \right)\left(d\phi\wedge \theta\delta(d\theta)\right)= d \left(\iota_{(\vec \epsilon)}{\cal J}^{(1|1)}\right)
\end{equation}
where $\ell_{\vec \epsilon}$ is the Lie derivative in superspace. Equation (\ref{susytraJ}) proves the requirement 4.

Next we note that in the case of the bosonic entropy current $J = d \phi$, we can construct an infinite
number of  currents by multiplying it by any function $f(\phi)$, since then $J_f = f(\phi) d\phi$  is clearly closed. It is immediate to verify
that the same construction can be extended to the entropy supercurrent. Indeed, introducing an arbitrary superfield $f(\phi,\theta)$ and defining
$ {\cal J}^{(1|1)}_f = f(\phi,\theta)  {\cal J}^{(1|1)}$,  by exterior differentiation it immediately follows:

\begin{equation}\label{superinf}
 d {\cal J}^{(1|1)}_f= d\, \left(f(\phi,\theta)  {\cal J}^{(1|1)}\right)=df(\phi,\theta)\wedge {\cal J}^{(1|1)}= 0.
\end{equation}
 However, in our supersymmetric setting, we can still construct another infinite set of conserved
 currents.

Let us introduce the following expression
\begin{equation}
\eta_{(0|1)}=( \theta \, \delta (d\theta) + \Pi \wedge \delta'(d\theta) )\,.
\end{equation}
$\eta_{(0|1)}$ is a $(0|1)$-form since the first term is a pure Dirac delta function (which carries no form degree) and the
second term is made of a 1-form, namely $\Pi$, and a $(-1)$-form, namely $\delta'(d\theta)$. Acting on $\eta_{(0|1)}$ with the
differential $d$, we have
\begin{eqnarray}\label{EC3B}
d \eta_{(0|1)} &=& d \Big( \theta \, \delta (d\theta) +  \Pi \wedge \delta'(d\theta) \Big) = d\Pi \wedge \delta'(d\theta) \nonumber \\
&=& d\theta\wedge d\theta \wedge \delta'(d\theta) =  - d\theta \wedge \delta(d\theta) = 0
\end{eqnarray}
In addition, we can define a new current
\begin{equation}\label{EC3C}
\eta_{(-1|1)} = \left(\theta \, \delta' (d\theta) +  \Pi \wedge \delta''(d\theta) \right) \,.
\end{equation}
where $\delta''(d\theta)$ is the second derivative of the Dirac delta function. The quantity $\eta_{(-1|1)}$ is a $(-1)$
form because the derivatives on Dirac delta functions count as negative form number. Again, $d \eta_{(-1|1)} =0$
using the properties of delta functions. Proceeding in this way we can define an infinite set of currents of the form
\begin{equation}\label{EC3D}
\eta_{(-n|1)} = \left(\theta \, \delta^{(n)} (d\theta) +  \Pi \wedge \delta^{(n+1)}(d\theta) \right) \,,
\end{equation}
satisfying $d\eta_{(-n|1)} =0$.

\subsection{Supersymmetric Entropy Current in $d+1$ dimensions}\label{d+1entr} As for the action and the
equations of motion it is straightforward to generalize  the entropy current from two to $(d+1)$-dimensions.
 
In the same way as for the two-dimensional case, we show that ${\cal J}^{(d|m)}$, as introduced in \ref{EC4alter}, is
off-shell closed. Indeed recalling that $d\theta^{\alpha}\,\delta(d\theta^{\alpha})=0$,
 so that
\begin{equation}\label{otherform}
{{\cal J}^{(d|m)}\equiv \frac{1}{d!}\epsilon_{I_1 \dots I_d}  d\phi^{I_1}  \wedge \dots \wedge d\phi^{I_d} (\theta)^m\bigwedge_{\alpha=1}^m \delta(d\theta^{\alpha})}\,,
\end{equation}
 we have, neglecting an overall sign
\begin{equation}
d {\cal J}^{(d|m)} =  \frac{1}{d!} \,\epsilon_{I_1 I_2,\dots I_d}d\phi^{I_1}  \wedge \dots \wedge d\phi^{I_d}\wedge
\sum_{\beta=1}^m d\theta^\beta\prod_{\sigma\neq \beta}\theta^\sigma\bigwedge_{\alpha=1}^m \delta(d\theta^{\alpha})=0\,.\label{dJuguale0}
\end{equation}

Therefore the entropy current is a closed form. This in particular implies that the entropy $S$ is conserved.
Indeed, from eq. (\ref{SrelJ}), we have

\begin{equation}\label{entropycons}
  0=  \int_{{S\cal M}^{(d+1|m)}}d {\cal J}^{(d|m)}= \int_{\partial ({S\cal M})^{(d|m)}} {\cal J}^{(d|m)}=\int_{\partial ({\cal M})^{(d)}}{ J}^{(bos)}=S(t_2)-S(t_1)\,,
\end{equation}
where $\partial ({S\cal M})^{(d|m)}$ denotes the boundary of ${S\cal M}^{(d+1|m)}$, represented by two
space-like hypersurfaces $(S\Sigma)_{t_1}^{(d|m)}$, $(S\Sigma)_{t_2}^{(d|m)}$ at the times $t_1,\,t_2$. The
integration on the Grassmann volume element $\prod_{\alpha=m}^1\theta^\alpha\bigwedge_{\beta=1}^m \delta(d\theta^{\beta})$ is
performed using the property (\ref{theproperty}).
Equation (\ref{entropycons}) thus implies the entropy conservation.

Finally we consider the transformation law of ${\cal J}^{(d|m)}$ under supersymmetry. We have

\begin{align}\label{unterderlinden}
\delta_{\vec \epsilon}{\cal J}^{(d|m)}&\equiv \,\ell_{\vec \epsilon}{\cal J}^{(d|m)}=
  (\iota_{\vec \epsilon}d+d\,\iota_{\vec \epsilon} ) \left[d\phi^{I_1}\wedge \dots \wedge d\phi^{I_d}
  (\theta)^m\bigwedge_{\alpha=1}^m \delta(d\theta^{\alpha})\right]\\ \nonumber &=d\, \left[\iota_{\vec \epsilon}\left(
  d\phi^{I_1}\wedge \dots \wedge d\phi^{I_d} (\theta)^m\bigwedge_{\alpha=1}^m \delta(d\theta^{\alpha})\right)\right]=
   d\left(\,\iota_{\vec \epsilon}{\cal J}^{(d|m)}\right)\,.
\end{align}
Integrating eq. (\ref{unterderlinden}) over a spatial
$(d|m)$ hypersurface $(S\Sigma)^{(d|m)}$ at a given time $t$  and using (\ref{SrelJ}), (\ref{unterderlinden}):
\begin{equation}
\delta_{\vec \epsilon} S=\int_{(S\Sigma)^{(d|m)}} \delta_{\vec \epsilon}{\cal J}^{(d|m)}=\int_{(S\Sigma)^{(d|m)}}  d\left(\,\iota_{\vec \epsilon}{\cal J}^{(d|m)}\right)=
\int_{\partial (S\Sigma)^{(d-1|m)}} \iota_{\vec \epsilon}{\cal J}^{(d|m)}=0\,,
\end{equation}
the last integral being zero since the boundary $\partial (S\Sigma)^{(d-1|m)}$ of $(S\Sigma)^{(d|m)}$ is located
at spatial infinity, where we assume all fields to vanish together with their derivatives.

 Actually, as in the bosonic case, there is an infinity of $d$-form
currents which are off-shell closed (and therefore their Hodge-dual are conserved). Indeed if we define

\begin{equation}\label{infinity1}
 {\cal J}^{(d|m)}_f =f(\phi^I,\theta)\, {\cal J}^{(d|m)}\,,
\end{equation}
then

\begin{equation}\label{extinf}
    d\,\left[f(\phi^I,\theta)\, {\cal J}^{(d|m)}\right]= \left[\frac{\partial f}{\partial\phi^J}d\phi^J+
    \frac{\partial f}{\partial\theta^\alpha}d\theta^\alpha\right]\,{\cal J}^{(d|m)}=0
\end{equation}


Finally, we show that we can define an infinite set of fermionic closed $(-n|1)$-superforms analogous to those
defined in the two-dimensional case (see eq.s (\ref{EC3B}), (\ref{EC3C}), (\ref{EC3D})). They are defined as

\begin{equation}
J^\alpha_{\beta_1\dots \beta_n}=\mathbb{P}^{\delta_1\dots
\delta_n}_{\beta_1\dots \beta_n}\,\hat{J}{}^\alpha_{\delta_1\dots
\delta_n}\,,\label{JP}
\end{equation}
where $\mathbb{P}$ is the projector onto the \emph{irreducible} $n$-fold symmetric product of the spinorial
representation and
\begin{align}
\hat{J}^\alpha_{\beta_1\dots
\beta_n}&\equiv\theta^\alpha\,\partial_{\beta_1}\cdots
\partial_{\beta_n}\prod_\beta \delta(d\theta^\beta)+\frac{i}{
d}\,\Pi_I \,(\Gamma^IC^{-1})^{\alpha\gamma}\partial_\gamma
\partial_{\beta_1}\cdots \partial_{\beta_n}\prod_\beta
\delta(d\theta^\beta)\,,\nonumber\\\partial_\beta&\equiv
\frac{\partial}{\partial d\theta^\beta}\,,\label{hatj}
\end{align}
where $C$ is the charge-conjugation matrix. The fermionic currents satisfy the conservation equation

\begin{equation}\label{fermioniccons}
   d\left( \mathbb{P}^{\delta_1\dots
\delta_n}_{\beta_1\dots \beta_n}\,\hat{J}^\alpha_{\delta_1\dots
\delta_n}\right) =0
\end{equation}

The proof is given in Appendix \ref{prooffermcons}.

\subsection{Generalized Expression for the Entropy Current}

In the present section, we consider a generalized form of the entropy current and its relation with the one given above. This is related to the fact that the integral forms can be seen also from a gauge fixing
point of view.\footnote{ This was the original point of view for introducing the PCO (Picture Changing Operator)
which are written in terms of integral form in superstring formulation. To make a long story short, we recall
that in the case of superstring the gauge symmetry is a local symmetry plus worldsheet diffeomorphisms and
therefore its quantization proceeds by fixing those symmetries by a gauge-fixing-BRST methods. In that process,
we have to choose a background metric and a background gravitino. For example, one simple choice is to set the
gravitino to zero. However, the corresponding ghost -- needed to implement the BRST formalism for that gauge
summetry -- is a commuting ghost (usually denoted by $\beta$) and the functional integral on it yields the Dirac
delta function for the gravitino. Obviously, one can choose a different gauge fixing.
\cite{Friedan:1985ge,Berkovits:2004px,Grassi:2004tv}.}

In our case, for the 1+1 dimensional case  we can consider the following expressions
\begin{equation}\label{GE1}
P = d \Phi + \Theta d \Theta\,, \quad \delta (d \Theta)\,, \quad d \Theta\,.
\end{equation}
where $\Phi$ and $\Theta$ are functions of the coordinates $\phi, \theta$. Therefore the generalised expression
becomes
\begin{equation}\label{GE2}
\tilde{{\cal J}}^{(1|1)} = P\wedge \Theta\,\delta( d \Theta)\,.
\end{equation}
It is straightforward to connect it to the original formula (\ref{EC1}) by expressing ({\ref{GE1}}) in terms of the
coordinates $\phi, \theta$. This can be easily done by observing
\begin{equation}\label{GE3}
\tilde{{\cal J}}^{(1|1)} = \left[ d\phi \Big( \partial_\phi \Phi + \Theta \partial_\phi \Theta \Big) + d \theta \Big( \partial_\theta \Phi
 + \Theta \partial_\theta
\Theta \Big) \right] \Theta\delta\Big( d \theta \partial_\theta \Theta + d\phi \partial_\phi \Theta \Big) =
\end{equation}
$$
=\left[ d\phi \, \partial_\phi \Phi +   d \theta  \,\partial_\theta \Phi  \right] \frac{\Theta}{\partial_\theta \Theta}
\left[ \delta\Big( d \theta  + d\phi \frac{1}{\partial_\theta \Theta} \partial_\phi \Theta \Big) \right].
$$
Using now
$$
\Theta=\partial_\theta \Theta\,\theta \equiv f(\phi)\,\theta;\quad\quad \partial_\phi \Theta=\partial_\phi f\,\theta
$$
we obtain
$$
\tilde{{\cal J}}^{(1|1)}=\left[ d\phi \partial_\phi \Phi + d \theta \partial_\theta \Phi  \right] \theta\left[ \delta (d \theta)  + d\phi \frac{1}{\partial_\theta \Theta} \partial_\phi \Theta \,
\partial_\phi f\theta\delta'(d\theta) \right]= d\phi\partial_\phi \,\Phi\wedge \theta\delta( d\theta).
$$

 This proves that the relation between the two formulas is simply the  determinant of
the (bosonic) Jacobian matrix ${\bf J}$ which, in the $d=1$ case, is just ${\bf J}=\partial_\phi \Phi$.\par
The generalization of the above derivation to the $(d+1)$-dimensional case is straightforward, though more involved.
Consider the following super-reparametrization:
\begin{equation}
\phi^I,\,\theta^\alpha\,\,\longrightarrow\,\,\,\Phi^I(\phi^J,\theta^\beta),\,\Theta^\alpha(\phi^J,\theta^\beta)\,,
\end{equation}
and the corresponding generalized entropy current:
\begin{equation}
\tilde{{\cal J}}^{(d|m)}=\frac{1}{d!}\epsilon_{I_1\dots I_d}\,P^{I_1}\wedge\dots P^{I_d}\left(\prod_{\beta=m}^1\Theta^\beta\right)\bigwedge_{\alpha=1}^m\delta(d\Theta^\alpha)\,,\label{GE22}
\end{equation}
where
\begin{equation}
P^I=d\Phi^I+\frac{i}{2}\bar{\Theta}\Gamma^I\,d\Theta\,.\label{PPP}
\end{equation}
Following the derivation in Appendix \ref{AppendixJacob} one finds:
\begin{equation}
\tilde{{\cal J}}^{(d|m)}={\rm det}\left({\bf J}\right)\,\frac{1}{d!}\epsilon_{I_1\dots I_d}\,\Pi^{I_1}\wedge\dots \Pi^{I_d}\left(\prod_{\beta=m}^1\theta^\beta\right)\bigwedge_{\alpha=1}^m\delta(d\theta^\alpha)={\rm det}\left({\bf J}\right)\,{{\cal J}}^{(d|m)}\,,\label{trapropJ}
\end{equation}
where ${\bf J}=({\bf J}_I{}^J)=(\frac{\partial \Phi^J}{\partial \phi^I})$ is the bosonic block of the super-Jacobian at $\theta^\alpha=0$.

The generalized expression (\ref{GE2}) (or (\ref{GE22})) could be the appropriate form in order to study the supersymmetric
entropy current in a fluid/gravity context (see for example
\cite{Bhattacharyya:2008jc},\cite{Bhattacharya:2012zx},\cite{Compere:2012mt}).  In the bosonic case the entropy
density is proportional to the black-hole horizon area. Then the supersymmetric version of the area increase
theorem would not change and   be still expressed as the statement
\begin{equation}\label{GE4}
\frac{ \partial}{\partial \lambda }  {\rm det {\bf J}} \geq 0\,,
\end{equation}
where $\lambda$ is the additional bosonic coordinate orthogonal to the super surface whose volume element is the
$(1|1)$ integral form  ({\ref{GE1}}).

Let us  check how the supersymmetry transformations act on this generalized expression for the entropy current.
Upon a supersymemtry transformation, using eq.  (\ref{trapropJ}) we find (up to an overall sign):
\begin{align}\label{GE5}
\delta_{\vec\epsilon} \tilde{{\cal J}}^{(d|m)} =&\delta_{\vec\epsilon} \left({\rm det}({\bf J}(\phi))\right)\,{{\cal J}}^{(d|m)}+{\rm det}({\bf J}(\phi))\delta_{\vec\epsilon} \tilde{{\cal J}}^{(d|m)}=\frac{\partial}{\partial \phi^K}{\rm det}({\bf J})\,\delta_{\vec\epsilon} \phi^K+\nonumber\\&+{\rm det}({\bf J})\,d\left(\iota_{\vec\epsilon} {{\cal J}}^{(d|m)}\right)= {\rm det}({\bf J})\,d\left(\iota_{\vec\epsilon} {{\cal J}}^{(d|m)}\right)\,,
\end{align}
where we have used the property that $\delta_{\vec\epsilon} \phi^K\propto \bar{\theta}\Gamma^K\epsilon$  is annihilated when multiplied by ${{\cal J}}^{(d|m)}$ due to the presence of the factor $\prod_\alpha \theta^\alpha$ in the latter. From eq. (\ref{GE5}) we see that, by supersymmetry, $\tilde{{\cal J}}^{(d|m)}$ still transforms by a total derivative only if the relation between $\Phi^I$ and $\phi^I$ is a \emph{volume preserving diffeomorphism}, i.e. if ${\rm det}({\bf J})=1$.

\subsection{Entropy Current for Supergravity}\label{sugra}

The  generalization of the entropy current of a fluid to the case of local supersymmetry can be given but is
somewhat problematic. First of all, it is not interesting from the point of view of the fluid/gravity
correspondence, where the holographic entropy corresponding to the  horizon-area of the boosted black-hole
solution is instead in correspondence with the entropy of the fluid at the boundary. However, we expect an
entropy current in supergravity to exist. It should reduce, in the rigid limit, to the supersymmetric entropy
current given in the previous sections and, in analogy to what happens in the rigid case, it should be
(Lorentz)-covariantly closed and should transform as a total covariant derivative under local supersymmetry.

 Let us work, for the sake of simplicity, in the case of $\mathcal{N}=1$, D=4 supergravity coupled to a set of chiral
multiplets $(z^i,\chi^i)$ together with their hermitian conjugates $(z^{\bar \imath}, \chi^{\bar\imath})$ ($i, \bar\imath=1,\cdots n_c$).
 Let $\Pi^a$ be the (bosonic) vielbein ($a=0,1,2,3$ denote anholonomic space-time indices) and $\Psi^\alpha$ the gravitino (fermionic)
one-form in superspace. In terms of a holonomic basis (spanned by ($d\phi^\mu,d\theta^\alpha$)), the anholomic
basis of 1-forms in superspace is defined by:
\begin{eqnarray}
\Pi^a&=& \Pi^a_\mu d\phi^\mu + \Pi^a_\alpha d\theta^\alpha\label{SVb}\\
\Psi^\alpha&=& \Psi^\alpha_\mu d\phi^\mu + \Psi^\alpha_\beta d\theta^\beta\,.\label{SVf}
\end{eqnarray}
We use the notation $a,b,c,\dots$ for Lorentz vector  indices, $\mu,\nu,\cdots$ for curved vector indices, while
Greek indices $\alpha, \beta, \gamma, \dots$ will denote the
 spinor-component  indices of the gravitino 1-form, running from 1 to 4. Like in the rigid supersymmetry case,  $I,J,K, \dots$
will denote 3-dimensional spatial vector indices.

The supersymmetry transformation laws under a superdiffeomorphism $\theta^\alpha \to \theta^\alpha +
\epsilon^\alpha(\phi,\theta)$ are:
\begin{eqnarray}\label{SG4}
\delta_\epsilon \Pi^a &=& i \bar \epsilon \Gamma^a \Psi\,,  \label{susylaw1}\\
\delta_\epsilon \Psi&=& \nabla \epsilon + L_a\,\Gamma^{ab}\epsilon   \Pi_b +\left[(
\Re S)+{\rm i} \gamma^5(\Im S)\right] \Gamma ^a  \epsilon  \Pi_a\,, \label{susylaw2}
\end{eqnarray}
where $ \mathrm{L}_a=\frac{i}{8} \chi^i \Gamma^{ab}\chi^{\bar \imath} g_{i\bar \jmath}$ is a current of
spin-$\frac12$  left-handed and right-handed chiral fields $\chi^i,\chi^{\bar \imath}$ respectively, $g_{i\bar
\jmath}$ is the Kaehler metric of the scalar-fields $\sigma$-model and $S(z^i,z^{\bar \imath})\equiv W(z)
e^{\frac{K}{2}}$ is the gravitino mass, $W$ being the superpotential and $K(z^i,z^{\bar \imath})$ the Kaehler
potential.

The most natural extension to supergravity  of eq. (\ref{EC4alter}), which has the property of reducing to
(\ref{EC4alter}) in the rigid limit, is given by the following expression:
\begin{equation}\label{NEC5}
{\cal J}_{(3|4)} =\frac{1}{3!} \,\epsilon_{I_1 \dots I_3} \Pi^{I_1}  \wedge \Pi^{I_2} \wedge \Pi^{I_3}\prod_{\alpha =4}^{1} \theta^\alpha
\bigwedge_{\beta=1}^4\delta(d\theta^{ \beta}) \,.
\end{equation}
Here we used as fermionic coordinates the same set used in rigid superspace (``free falling" frame in the
fermionic directions of superspace). To make this expression manifestly covariant in superspace,  we can
equivalently rewrite it as:
\begin{equation}\label{NEC5b}
\boxed{{\cal J}_{(3|4)} =\frac{1}{3!} \,\epsilon_{I_1 \dots I_3} \Pi^{I_1}  \wedge \Pi^{I_2} \wedge \Pi^{I_3} \prod_{\alpha =4}^{1} \xi^\alpha
\bigwedge_{\beta=1}^4\delta(\Psi^{ \beta})}
\end{equation}
where we have introduced a set of spinors $\xi^\alpha$ in superspace, defined by $\xi^\alpha(x,\theta) \equiv \Psi^\alpha_\beta \,\theta^\beta$ (see eq. (\ref{SVf})).
Note that ${\cal J}_{(3|4)}$ is a 3-form with picture-number 4. We now show that it is (covariantly)
closed. Indeed
\begin{eqnarray}\label{OEC6}
\nabla {\cal J}_{(3|4)} &=&  \frac{\rm i}{4} \,\epsilon_{I_1 I_2 I_3}\, \bar\Psi\wedge \Gamma^{I_1}
\Psi  \wedge \Pi^{I_2} \wedge \Pi^{I_3}\prod_{\alpha =4}^{1} \xi^\alpha
\bigwedge_{\beta=1}^4 \delta(\Psi^{\beta})  \nonumber + \\
&+& \frac{2}{3} \, \epsilon_{I_1 I_2 I_3}\Pi^{I_1}  \wedge \Pi^{I_2} \wedge \Pi^{I_3}\prod_{\alpha =4}^{1} \xi^\alpha\wedge \left[ \sum_{\beta=1}^4
 \delta'(\Psi^{\beta})\wedge \nabla\Psi^{\beta}\bigwedge_{\alpha\neq \beta}
\delta(\Psi^{\alpha})\right]\nonumber + \\
&+& \frac{2}{3} \, \epsilon_{I_1 I_2 I_3}\Pi^{I_1}  \wedge \Pi^{I_2} \wedge \Pi^{I_3}\wedge\left[ \sum_{\alpha=1}^4
  \nabla\xi^{\alpha}\prod_{\gamma\neq \alpha=4}^1
 \xi^{\gamma}\right]\,\bigwedge_{\beta=1}^4 \delta(\Psi^{\beta})
\end{eqnarray}
where we  used the torsion constraint in superspace, discussed in (\ref{SG2}) in the Appendix, implying $\nabla
\Pi^I= \mathcal{D}\Pi^I=d \Pi^I - \omega^I_a \wedge \Pi^a = \frac{\rm i}2\,{\bar\Psi}  \Gamma^I \Psi $, where
$\mathcal{D}$ denotes Lorentz-covariant derivative.

Because of the presence of the current ${\bar\Psi} \Gamma^I \Psi $, the first line of eq. (\ref{OEC6}) actually
vanishes  in force  of the identity $\Psi^\alpha \delta(\Psi^\alpha) = 0$. As far as the second and third  terms
are concerned, to directly show that they sum to zero is a bit involved, however they are easily shown to vanish
by using, instead of (\ref{NEC5b}), the equivalent expression (\ref{NEC5}) for the entropy current, since in
this formulation we can make use of the relation $d\theta^\alpha \,\delta(d\theta^\alpha)=0$.

    In conclusion:
    \begin{equation}\label{prom}
       \nabla {\cal J}_{(3|4)}=0\,.
    \end{equation}

Moreover, it is easy to show that ${\cal J}_{(3|4)}$ is invariant under supersymmetry transformations, up to a total
covariant derivative. To this purpose, we can use the general property that the supersymmetry transformation is
actually a Lie derivative in superspace along the fermionic tangent vectors dual to the gravitini $\Psi$.
Moreover, if the Lie derivative acts on Lorentz-covariant forms, say $\omega$, we may replace the ordinary
differential with Lorentz-covariant differentials:
\begin{eqnarray}
\delta_\epsilon \omega &=&  \ell_\epsilon \omega = \imath_\epsilon d\omega + d(\imath_\epsilon \omega )\nonumber\\
&=&\imath_\epsilon \nabla \omega + \nabla (\imath_\epsilon \omega )\,.
\end{eqnarray}
We then have:
\begin{eqnarray}
\delta_\epsilon {\cal J}_{(3|4)}&=&   \imath_\epsilon \nabla {\cal J}_{(3|4)} + \nabla (\imath_\epsilon {\cal J}_{(3|4)})=\nabla (\imath_\epsilon {\cal J}_{(3|4)}) \,.
\end{eqnarray}

The above definition (\ref{NEC5b}) of the entropy current, which is a direct extension of the definition in rigid superspace (\ref{EC4alter}),
is not completely satisfying since it does not account for the dynamics of the gravitino fields. An alternative
definition, which encodes a non trivial dynamics of superspace, is the one  generalizing to local supersymmetry
the definition (\ref{EC4}), that is:
\begin{equation}
\label{altersugra}
{J^{(d|m)} = \frac{1}{d!}\epsilon_{I_1 I_2 I_3}  \Pi^{I_1}  \wedge \dots \wedge
 \Pi^{I_3} \bigwedge_{\alpha=1}^m \delta(\Psi^{\alpha})}=J^{(bos)}\,\bigwedge_{\alpha=1}^m
\delta(\Psi^{\alpha})\,.
\end{equation}
This expression is covariantly closed (a proof will be given in Appendix \ref{sugraproof}). Its physical
relation to the bosonic entropy current requires further study. We just observe that the above form is related to its rigid counterpart (\ref{EC4}) through the \emph{super-determinant} of the supervielbein matrix
 \begin{equation}\label{nonsochelabelmettere}
{{J}^{(d|m)} ={\rm sdet}(E^M{}_N) \frac{1}{d!}\epsilon_{I_1 \dots I_d}  d\phi^{I_1}  \wedge \dots \wedge
d \phi^{I_d} \bigwedge_{\alpha=1}^m \delta(d\theta^{\alpha})}\,,
\end{equation}
where
\begin{equation}
E^M{}_N=\left(\begin{matrix}\Pi^I{}_J & \Pi^I{}_\alpha\cr   \Psi^\beta{}_J & \Psi^\beta{}_\alpha\end{matrix}\right)\,.
\end{equation}
We refer the reader to Appendix \ref{AppendixJacob} for a proof of the above statement.

\section{ Conclusions}

In this paper we have developed a supersymmetric extension of the non dissipative fluid dynamics using as a
starting point the effective theory approach of \cite{Dubovsky:2011sj}. By  introducing the supersymmetric
partners of the comoving coordinate fields and suitably extending the entropy density and chemical potential
fields to superfields, we have shown that the action functional is invariant under supersymmetry
transformations. In this setting the bosonic entropy density turns out to be the lowest component of the
supersymmetric extension. A drawback of this result is that the natural supersymmtric extension of the current
density d-form
$
J^{(d|m)} = \frac{1}{d!}\epsilon_{I_1 \dots I_d}  \Pi^{I_1}  \wedge \dots \wedge \Pi^{I_d}
$
is not closed. We have then developed an alternative approach to the supersymmetric entropy current based on the
formalism of the integral forms. In this new setting the new current is identically closed as it happens in the
bosonic case and the Berezin integration in superspace  reproduces the bosonic entropy formula. Moreover in this
new setting an extension of the entropy current to supergravity seems feasible and we have explicity
constructed it in the case of $\mathcal{N}=1$ supergravity coupled to chiral multiplets.\\
It would be interesting to see what is the relation, if any, between the two approaches and to explicitly work
out the physical implications in both cases. This is left to future investigation.

\section{Acknowledgements}
This work
was partially supported by the Italian MIUR-PRIN contract 2009KHZKRX-007 ``Symmetries of the
Universe and of the Fundamental Interactions''. We would like to thank V. Penna and  G. Policastro for useful
discussions.

\appendix

\section{Appendix}
\subsection{Conventions}\label{conventions0}
We generally use   $\mu,\nu,\cdots=0,1,\cdots,d$ to denote space-time indices and $\alpha,\beta,\cdots=
1,\cdots,m$ to denote spinor indices. We adopt the ``mostly plus'' signature of the metric and the following
definition of the Hodge dual operation in $D=d+1$ space-time dimensions:
\begin{equation}
{}^\star (dx^{\mu_1}\wedge\dots\wedge dx^{\mu_k})=\frac{1}{(D-k)! \sqrt{|g|}}\,\epsilon^{\mu_1\dots \mu_k \mu_{k+1}\dots\mu_{D}}\,dx_{\mu_{k+1}}
\wedge\dots\wedge dx_{\mu_{D}}\,.
\end{equation}
In this convention we have:
\begin{align}
{}^\star {}^\star \omega^{(p)}&=(-1)^{p(D-p)+1}\,\omega^{(p)}\,\,\,\,,\,\,\,\
\omega^{(p)}\wedge{}^\star \eta^{(p)}=-\frac{\sqrt{|g|}}{p!}\,\omega_{\mu_1\dots \mu_p}\,\eta^{\mu_1\dots \mu_p}\, d^D x\,,
\end{align}
where we have used, for a generic $p$--form $\omega^{(p)}$, the following representation:
\begin{equation}
\omega^{(p)}=\frac{1}{p!}\,\omega_{\mu_1\dots \mu_p}\,dx^{\mu_1}\wedge\dots\wedge dx^{\mu_p}\,.
\end{equation}
We also use the convention:
\begin{equation}
dx^{\mu_1}\wedge\dots\wedge dx^{\mu_D}=\epsilon^{\mu_1\dots \mu_D}\, d^Dx\,.
\end{equation}

\subsection{Properties of the Integral Forms}\label{A2}

In this section we briefly recall the definition of
``integral forms" and their main properties referring mainly to \cite{mare} for
a detailed exposition.

The problem is that we can build the space $\Omega^k$ of $k$-superforms out of basic 1-superforms
$d\theta^{\alpha}$ and their wedge products, however these products are necessarily commutative, since the
$\theta_\alpha$'s are odd variables. Therefore, together
 with the differential operator $d$, the spaces $\Omega^k$ form a differential complex
\begin{equation}
0\overset{d}{\longrightarrow}\Omega^{0}\overset{d}{\longrightarrow}%
\Omega^{1}\dots\overset{d}{\longrightarrow}\Omega^{n}\overset
{d}{\longrightarrow}\dots\label{comA}%
\end{equation}
which is bounded from below, but not from above.

One simple way to define the ``integral forms'' is to introduce a new sheaf containing, among other object to be
defined, new basic forms $\delta (d\theta)$. We think of $\delta(d\theta)$ as an operator acting formally on the
space of superforms as the usual Dirac's delta measure. We write this as
\[ \left\langle f(d\theta) ,\delta(d\theta) \right\rangle = f(0), \]
where $f$ is a superform. Moreover we consider more general objects such as
the derivatives $\delta^{(n)}(d\theta)$. Here we have
\[ \left\langle f(d\theta) ,\delta^{(n)}(d\theta) \right\rangle =
- \left\langle f'(d\theta), \delta^{(n-1)}(d\theta) \right\rangle = (-1)^n f^{(n)}(0), \]
like the usual Dirac $\delta$ measure. Moreover we can consider objects
such as $g(d\theta) \delta(d\theta)$, which act by first multiplying by $g$ then applying $\delta(d\theta)$ (in analogy with a measure of type $g(x) \delta(x)$), and so on. The formal properties above imply in addition some simple relations:
\begin{equation}
\delta(d\theta)\wedge \delta(d\theta^{\prime})=-\delta(d\theta^{\prime}%
)\wedge \delta(d\theta), \ d\theta\wedge \delta(d\theta)=0, \ d\theta\wedge \delta^{\prime}%
(d\theta)=-\delta(d\theta).\label{comB}%
\end{equation}

The systematic exposition of these rules can be found in \cite{integ}. An interesting consequence of this
procedure is the existence of ``negative degree'' forms, which are those which reduce the degree of forms (e.g.
$\delta'(d\theta)$ has degree $-1$).

We introduce also the \textit{picture number} by counting the number of delta
functions (and their derivatives) and we denote by $\Omega^{r|s}$ the
$r$-forms with picture $s$. For example the integral form
\begin{equation}
dx^{[\mu_{1}}\wedge \dots {}\wedge dx^{\mu_{l}]}\wedge d\theta^{(\alpha_{1}}\wedge \dots {}\wedge d\theta^{\alpha_{r})}\wedge
\delta(d\theta^{[\alpha_{r+1}})\wedge \dots{}\wedge \delta(d\theta^{\alpha_{r+s}]})\ \label{comC}%
\end{equation}
is an $(r+l)$-from with picture number $s$. All indices $\mu_{i}$ and $\alpha_{r+1}, \dots \alpha_{r+s}$ are
antisymmetrized while $\alpha_{1}, \dots \alpha_{r}$ are symmetrized. Indeed, by also adding derivatives of
delta forms $\delta^{(n)}(d\theta)$, even negative form-degree can be considered, e.g. a form of the type:
\begin{equation}
\delta^{(n_{1})}(d\theta^{\alpha_{1}})\wedge \dots{}\wedge \delta^{(n_{s})}(d\theta^{\alpha_{s}})
\label{comCA}%
\end{equation}
is a $-(n_{1}+\dots n_{s})$-form with picture $s$. Clearly $\Omega^{k | 0}$ is
just the set $\Omega^k$ of superforms, for $k \geq 0$.

We can formally expand the Dirac delta functions in series
\begin{equation}
\delta\left(  d\theta^{1}+d\theta^{2}\right)  =\sum_{j}\frac{\left(
d\theta^{2}\right)  ^{j}}{j!}\delta^{(j)}(d\theta^{1})
\end{equation}
Recall that any usual superform is a polynomial in the $d\theta,$ therefore only a
finite number of terms really matter in the above sum, when we apply it to a superform. Infact, applying the formulae above, we have for example,
\begin{equation}
\left\langle (d\theta^{1})^{k},\sum_{j}\frac{\left(  d\theta^{2}\right)  ^{j}%
}{j!}\delta^{(j)}(d\theta^{1})\right\rangle =(-1)^{k}(d\theta^{2})^{k}%
\end{equation}
Notice that this is equivalent to the effect of replacing $d\theta^{1}$ with
$-d\theta^{2}.$ We could have also interchanged the role of $\theta^{1}$ and
$\theta^{2}$ and  the result would be to replace $d\theta^{2}$ with
$-d\theta^{1}.$ Both procedures correspond precisely to the action we expect
when we apply the $\delta\left(  d\theta^{1}+d\theta^{2}\right)  $ Dirac
measure.

The integral forms form a new complex as follows
\begin{equation}
\dots\overset{d}{\longrightarrow}\Omega^{(r|q)}\overset{d}{\longrightarrow
}\Omega^{(r+1|q)}\dots\overset{d}{\longrightarrow}\Omega^{(p+1|q)}\overset
{d}{\longrightarrow}0 \label{comD}%
\end{equation}
where $\Omega^{(p+1|q)}$ is the top ``form'' $dx^{[\mu_{1}}\wedge \dots {}\wedge dx^{\mu_{p+1}%
]}\wedge \delta(d\theta^{\lbrack \alpha_{1}})\wedge \dots{}\wedge \delta(d\theta^{\alpha_{q}]})$ which can be
integrated on the supermanifold. As in the usual commuting geometry, there is an isomorphism between the
cohomologies $H^{(0|0)}$ and $H^{(p+1|q)}$ on a supermanifold of dimension $(p+1|q)$. In addition, one can
define two operations acting on the cohomology groups $H^{(r|s)}$ which change the picture number $s$ (see
\cite{integ}).

Given a function $f(x,\theta)$ on the superspace, we
define its integral by the super top-form $\omega^{(p+1|q)}=f(x,\theta
)d^{p+1}x\delta(d\theta^{1})\dots\delta(d\theta^{q})$ belonging to
$\Omega^{(p+1|q)}$ as follows
\begin{equation}
\int_{\mathbb{R}^{(p+1|q)}}\omega^{(p+1|q)}=\frac{1}{q!}\,\epsilon^{\alpha_{1}\dots \alpha_{q}%
}\partial_{\theta^{\alpha_{1}}}\dots\partial_{\theta^{\alpha_{q}}}\int_{\mathbb{R}%
^{p+1}}f(x,\theta) d^{p+1}x \label{comE}%
\end{equation}
where the last equality is obtained by integrating on the delta functions and selecting the bosonic top form.
The remaining integrals are the usual integral of densities and the Berezin integral. It is easy to show that
indeed the measure is invariant under general coordinate changes and the density transform as a Berezinian with
the superdeterminant (see Appendix \ref{AppendixJacob}). Note that in particular we have
\begin{equation}\label{inttetta}
  \int \theta^m\cdots\theta^1\bigwedge_{\beta=1}^m\delta(d\theta^\beta)=1\,.
\end{equation}

\subsection{Transformation Properties of $\tilde{\cal J}^{(d|m)}$}\label{AppendixJacob}
In this Appendix we wish to prove eq. (\ref{trapropJ}). To this end it is instructive to start from the  slightly more general situation of a current having the form:
\begin{equation}
\mathbb{J}^{(d|m)}=\frac{1}{d!}\epsilon_{I_1\dots I_d}\,P^{I_1}\wedge\dots P^{I_d}\left(\prod_{\beta=m}^1\Theta^\beta\right)\bigwedge_{\alpha=1}^m\delta(\Psi^\alpha)\,,\label{GE221}
\end{equation}
where:
\begin{equation}
P^I=P^I{}_J\,d\phi^J+P^I{}_\alpha\,d\theta^\alpha\,\,,\,\,\,\Psi^\alpha=\Psi^\alpha{}_I\,d\phi^I+\Psi^\alpha{}_\beta\,d\theta^\beta\,,
\end{equation}
are even and odd 1-forms, respectively, and $\Theta^\beta(\phi,\theta)$ are odd superfield 0-forms. Expanding $\Theta^\alpha$ in products of  $\theta$'s we find:
\begin{equation}
\Theta^\beta(\phi,\theta)=\Theta^\alpha{}_\beta(\phi)\,\theta^\beta+\Theta^\alpha{}_{\beta\gamma}(\phi)\,\theta^\beta \theta^\gamma+\dots
\end{equation}
We see that only the first term in the above expansion contributes to the $m$-fold product of the $\Theta$'s, so that:
\begin{equation}
\prod_{\beta=1}^m\Theta^\beta={\rm det}(\Theta^\alpha{}_\beta(\phi))\,\prod_{\beta=1}^m\theta^\beta\,.
\end{equation}
As far as the integral-form part of the current is concerned, let us use the following properties:
\begin{align}
\bigwedge_{\alpha=1}^m\delta(M^\alpha{}_\beta\,d\theta^\beta)&=\frac{1}{{\rm det}(M^\alpha{}_\beta)}\,\bigwedge_{\alpha=1}^m\delta(d\theta^\alpha)\,,\label{prop1}\\
\delta(d\theta^\alpha+F^\alpha{}_I d\phi^I)&=\sum_{k=0}^m\frac{1}{k!}\delta^{(k)}(d\theta^\alpha)\,(F^\alpha{}_{I} d\phi^{I})^k\,,\label{prop2}
\end{align}
where $\delta^{(k)}$ denotes the $k^{th}$-order derivative of the delta-function and $(F^\alpha{}_{I} d\phi^{I})^k$ the $k$-fold wedge product of the form $F^\alpha{}_{I} d\phi^{I}$.\par
Using (\ref{prop1}) and (\ref{prop2}) we can rewrite (\ref{GE221}) as follows:
\begin{align}
\mathbb{J}^{(d|m)}&=\frac{1}{d!}\epsilon_{I_1\dots I_d}\,P^{I_1}\wedge\dots P^{I_d}\frac{{\rm det}(\Theta^\alpha{}_\beta)}{{\rm det}(\Psi^\alpha{}_\beta)}(\theta)^m\,\bigwedge_{\alpha=1}^m\delta(d\theta^\alpha+F^\alpha{}_I d\phi^I)=\nonumber\\
&=\frac{1}{d!}\epsilon_{I_1\dots I_d}\,P^{I_1}\wedge\dots P^{I_d}\frac{{\rm det}(\Theta^\alpha{}_\beta)}{{\rm det}(\Psi^\alpha{}_\beta)}(\theta)^m\,\bigwedge_{\alpha=1}^m\sum_{k=0}^m\frac{1}{k!}\delta^{(k)}(d\theta^\alpha)\,(F^\alpha{}_{I} d\phi^{I})^k\,,
\end{align}
where $F^\alpha{}_I\equiv \Psi^{-1\,\alpha}{}_\beta\,\Psi^\beta{}_I$.
Next we use the fact that the presence of $\prod_{\beta=1}^m\theta^\beta$ singles out the order-0 terms in $\theta$ of all the other factors in the above expression. In particular, being $P^I{}_\alpha$ and $F^\alpha{}_I$ odd functions of $\phi,\,\theta$, they vanish when multiplied with  $\prod_{\beta=1}^m\theta^\beta$, so that:
\begin{align}
\mathbb{J}^{(d|m)}&={\rm det}(P^{(0)\,I}{}_J)\,\frac{{\rm det}(\Theta^\alpha{}_\beta)}{{\rm det}(\Psi^{(0)\,\alpha}{}_\beta)}\,\frac{1}{d!}\epsilon_{I_1\dots I_d}\,d\phi^{I_1}\wedge\dots d\phi^{I_d}(\theta)^m\,\bigwedge_{\alpha=1}^m\delta(d\theta^\alpha)\,,
\end{align}
where $P^{(0)\,I}{}_J(\phi)$ and $\Psi^{(0)\,\alpha}{}_\beta(\phi)$ denote the $\theta^\alpha=0$ components of the matrices
$P^{I}{}_J(\phi,\theta)$ and $\Psi^{\alpha}{}_\beta(\phi,\theta)$.\par
Consider now the particular case in which $\Psi^\alpha=d\Theta^\alpha$ and $P^I$ have the form in (\ref{PPP}). We have:
\begin{align}
P^{(0)\,I}{}_J&=\left.\frac{\partial \Phi^I}{\partial\phi^J}\right\vert_{\theta=0}={\bf J}_J{}^I\,\,\,;\,\,\,\,
\Psi^{(0)\,\alpha}{}_\beta=\left.\frac{\partial \Theta^\alpha}{\partial\theta^\beta}\right\vert_{\theta=0}=\Theta^\alpha{}_\beta\,.
\end{align}
The current $\mathbb{J}^{(d|m)}$ coincides with  $\tilde{J}^{(d|m)}$ and we find eq. (\ref{trapropJ}).\par
Inspired by the alternative definition of the entropy current in   (\ref{EC4}) and its supergravity version (\ref{altersugra}), let us repeat the above derivation for the following form:
\begin{equation}
\hat{\mathbb{J}}^{(d|m)}=\frac{1}{d!}\epsilon_{I_1\dots I_d}\,P^{I_1}\wedge\dots P^{I_d}\bigwedge_{\alpha=1}^m\delta(\Psi^\alpha)\,.\label{GE222}
\end{equation}
Using (\ref{prop1}) and (\ref{prop2}), we can write:
\begin{align}
\hat{\mathbb{J}}^{(d|m)}&=\frac{1}{d!}\epsilon_{I_1\dots I_d}\,P^{I_1}\wedge\dots P^{I_d}\frac{1}{{\rm det}(\Psi^\alpha{}_\beta)}\bigwedge_{\alpha=1}^m\sum_{k=0}^m\frac{1}{k!}\delta^{(k)}(d\theta^\alpha)\,(F^\alpha{}_{I} d\phi^{I})^k\,.
\end{align}
Notice, however, that now all the terms in the sum on the right hand side contribute, and each will select a  different term in the expansion of the product of the $P$'s. A careful derivation yields:
 \begin{align}
\hat{\mathbb{J}}^{(d|m)}&={\rm sdet}(E^M{}_N)\frac{1}{d!}\epsilon_{I_1\dots I_d}\,d\phi^{I_1}\wedge\dots d\phi^{I_d}\bigwedge_{\alpha=1}^m\delta(d\theta^\alpha)\,,
\end{align}
where the the indices $M,\,N$ run over the spatial and spinor ones $I,\,\alpha$, the matrix $E^M{}_N$ is defined as:
\begin{equation}
E^M{}_N=\left(\begin{matrix}P^I{}_J & P^I{}_\alpha\cr   \Psi^\beta{}_J & \Psi^\beta{}_\alpha\end{matrix}\right)\,,
\end{equation}
and ``sdet'' denotes the \emph{super-determinant}:
\begin{equation}
{\rm sdet}(E^M{}_N)=\frac{1}{{\rm det}(\Psi^\alpha{}_\beta)}\,{\rm det}\left(P^I{}_J-P^I{}_\beta\,\Psi^{-1\,\beta}{}_\alpha\,\Psi^\alpha{}_J\right)\,.
\end{equation}

\subsection{Proof of eq. (\ref{fermioniccons})}\label{prooffermcons}
We wish here to show that the form defined in (\ref{JP}) is closed. We start computing the exterior derivative of the current
\begin{align}
\hat{\cal J}^\alpha_{\beta_1\dots
\beta_n}&\equiv\theta^\alpha\,\partial_{\beta_1}\cdots
\partial_{\beta_n}\prod_\beta \delta(d\theta^\beta)+\frac{i}{
d}\,\Pi_I \,(\Gamma^IC^{-1})^{\alpha\gamma}\partial_\gamma
\partial_{\beta_1}\cdots \partial_{\beta_n}\prod_\beta
\delta(d\theta^\beta)\,,\nonumber\\\partial_\beta&\equiv
\frac{\partial}{\partial d\theta^\beta}\,,\label{hatjb}
\end{align}
where $C$ is the charge-conjugation matrix.
We shall need to use the properties:
\begin{align}
d\theta^\alpha\,\partial_{\beta_1}\cdots
\partial_{\beta_k}\prod_\beta
\delta(d\theta^\beta)&=k\,\delta^\alpha_{(\beta_1}\partial_{\beta_2}\cdots
\partial_{\beta_k)}\prod_\beta
\delta(d\theta^\beta)\,,\nonumber\\
d\theta^\alpha\,d\theta^\gamma\partial_{\beta_1}\cdots
\partial_{\beta_k}\prod_\beta
\delta(d\theta^\beta)&=k(k-1)\,\delta^{(\alpha\gamma)}_{(\beta_1\beta_2}\partial_{\beta_3}\cdots
\partial_{\beta_k)}\prod_\beta
\delta(d\theta^\beta)\,.
\end{align}
Exterior derivation of (\ref{hatjb}) then yields:
\begin{align}
d\hat{\cal J}^\alpha_{\beta_1\dots
\beta_n}&=n\,\delta^\alpha_{(\beta_1}\partial_{\beta_2}\cdots
\partial_{\beta_n)}\prod_\beta
\delta(d\theta^\beta)+\frac{i}{
d}\,\frac{i}{2}\,d\theta^TC\Gamma_Id\theta
\,(\Gamma^IC^{-1})^{\alpha\gamma}\partial_\gamma
\partial_{\beta_1}\cdots \partial_{\beta_n}\prod_\beta
\delta(d\theta^\beta)=\nonumber\\=&
n\,\delta^\alpha_{(\beta_1}\partial_{\beta_2}\cdots
\partial_{\beta_n)}\prod_\beta
\delta(d\theta^\beta)-\frac{n(n+1)}{
2d}\,(\Gamma^IC^{-1})^{\alpha\gamma}\,(C\Gamma_I)_{(\gamma\beta_1} \,
\partial_{\beta_2}\cdots \partial_{\beta_n)}\prod_\beta
\delta(d\theta^\beta)\nonumber\\=&n\,\delta^\alpha_{(\beta_1}\partial_{\beta_2}\cdots
\partial_{\beta_n)}\prod_\beta
\delta(d\theta^\beta)-\nonumber\\&- \frac{n}{
2d}\,(\Gamma^IC^{-1})^{\alpha\gamma}\,\left(2(C\Gamma_I)_{\gamma(\beta_1}
\,
\partial_{\beta_2}\cdots \partial_{\beta_n)}+(n-1)\,(C\Gamma_I)_{(\beta_1\beta_2}
\,
\partial_{\beta_3}\cdots \partial_{\beta_n)}\partial_\gamma\right)\prod_\beta
\delta(d\theta^\beta)=\nonumber\\
=&- \frac{n(n-1)}{
2d}\,(\Gamma^IC^{-1})^{\alpha\gamma}\,(C\Gamma_I)_{(\beta_1\beta_2}
\,
\partial_{\beta_3}\cdots \partial_{\beta_n)}\partial_\gamma\prod_\beta
\delta(d\theta^\beta)\,,
\end{align}
the latter term vanishes upon contraction with $\mathbb{P}$.

\subsection{Covariant Closure  of $ {J}^{(3|4)}$}\label{sugraproof}
In this Appendix we wish to prove that the current  $ {J}^{(3|4)}$ is covariantly closed. The supertorsion $T^a$ and the gravitino field-strength $\rho $ 2-forms in $N=1$ superspace are defined as
follows:
\begin{eqnarray}\label{SG1}
T^a &=& d V^a -\omega^a_{~b} \wedge V^b - \frac{i}{2}\, \bar\Psi \Gamma^a \Psi\,,  \\
\rho &\equiv& \nabla \Psi = d \Psi - \frac{1}{4} \Gamma^{ab} \omega_{ab} \Psi\,,
\end{eqnarray}
where $\omega^{ab}$ is the spin connection. Their on-shell parametrization in superspace (superspace
constraints) is
\begin{eqnarray}
T^a &=& 0 \label{SG2} \\
\rho &=&   \rho_{ab} V^a \wedge V^b + \mathrm{L}_a\, \Gamma_5 \, \Gamma^{ab}\Psi \wedge V^b +
\left[(\Re S) + {\rm i}\Gamma^5 (\Im S)\right]\Gamma^a\,\Psi\wedge V_a\,, \label{SG3}
\end{eqnarray}
where $\rho_{ab}V^a\wedge V^b$ is the supercovariant gravitino field strength,  $ \mathrm{L}_a=\frac{i}{8}
\chi^i \Gamma^{ab}\chi^{\bar \imath} g_{i\bar \jmath}$ is a current of spin-$\frac12$  left-handed and
right-handed chiral fields $\chi^i,\chi^{\bar \imath}$ respectively, $g_{i\bar \jmath}$ is the Kaehler metric of
the scalar-fields $\sigma$-model and $S(z^i,z^{\bar \imath})\equiv W(z) e^{\frac{K}{2}}$ is the gravitino mass,
$W$ being the superpotential and $K(z^i,z^{\bar \imath})$ the Kaehler potential. \footnote{Note that in this
formalism $\rho_{ab}V^a_\mu V^b_\nu$ is the supercovariant gravitino field-strength while $\rho_{\mu\nu}$ is the
ordinary field-strength $\nabla_{[\mu}\Psi_{\nu]}$.}

 We now show that the entropy current defined in (\ref{altersugra}) is (covariantly) closed. Indeed
\begin{eqnarray}\label{OEC6b}
\nabla {J}^{(3|4)} &=&  \frac{\rm i}{4} \,\epsilon_{I_1 I_2 I_3}\, \bar\Psi\wedge \Gamma^{I_1}
\Psi  \wedge V^{I_2} \wedge V^{I_3}
\bigwedge_{\beta=1}^4 \delta(\Psi^{\beta}) \nonumber + \\
&+& \frac{2}{3} \, \epsilon_{I_1 I_2 I_3}V^{I_1}  \wedge V^{I_2} \wedge V^{I_3}\wedge \left[ \sum_{\beta=1}^4
 \delta'(\Psi^{\beta})\wedge \nabla\Psi^{\beta}\bigwedge_{\alpha\neq \beta}
\delta(\Psi^{\alpha})\right]\,,
\end{eqnarray}
where we  used the constraint (\ref{SG2}), implying $  d V^I - \omega^I_a \wedge V^a = \frac{\rm i}2\,{\bar\Psi}
\Gamma^I \Psi $.

Because of the presence of the current ${\bar\Psi} \Gamma^I \Psi $, the first line of eq. (\ref{OEC6b}) actually
vanishes  in force  of the identity $\Psi^\alpha \delta(\Psi^\alpha) = 0$. As far as the second term is
concerned, we observe that substituting to $\nabla\Psi^{\beta}= \rho^\beta$ the right hand side of equation
(\ref{SG3}), we get four kinds of contributions: the terms with five vierbeine identically vanish in four
dimensions; as far as the contribution of the other three terms is concerned, they contain the products
$\Gamma_5 \Gamma^{ab}\,\Psi\wedge V_b$,
 $\Gamma^{a}\Psi\wedge V_a$ and  $\Gamma_5 \Gamma^{a}\Psi\wedge V_a$.
All of them, however, give a vanishing contribution owing to the traceless property of the $\Gamma$-matrix
algebra.
 Take for example the term
$$
-{\rm i}(\Im S)\Gamma ^a \Gamma^5 \Psi\wedge V_a.
$$
We have:
\begin{eqnarray}\label{psiV}
 &&   -{\rm i}(\Im S)\epsilon_{I_1 \dots I_3} V^{I_1}  \wedge V^{I_2} \wedge V^{I_3}  \wedge V_a  \left[ \sum_{\beta=1}^4
 \delta'(\Psi^{\beta})\wedge
    (\Gamma^a \,\Gamma^5)^\beta_{\,\,\gamma}\,
    \Psi^\gamma\bigwedge_{\alpha\neq \beta}
\delta(\Psi^{\alpha})\right]\nonumber\\
&=& {\rm i}(\Im S)\epsilon_{I_1 \dots I_3} V^{I_1}  \wedge V^{I_2} \wedge V^{I_3}  \wedge V_a  \left[ \sum_{\beta=1}^4
    (\Gamma^a \,\Gamma^5)^\beta_{\,\,\beta}\,
     \bigwedge_{\alpha }^4
\delta(\Psi^{\alpha})\right]= 0\,,
\end{eqnarray}
where we used the property $\delta'(\Psi^\gamma)\Psi^\gamma =-\delta(\Psi^\gamma)$.

    By the same argument one finds that also the terms proportional to    $ \Gamma^{a} \Psi$ and $\Gamma^5\Gamma^{ab} \Psi$
    give vanishing contribution since  $Tr (\Gamma^5\Gamma^{ab})= Tr (\Gamma^{a})=0$.

    In conclusion:
    \begin{equation}\label{promb}
       \nabla {J}^{(3|4)}=0\,.
    \end{equation}

\end{document}